\documentclass[manuscript]{aastex}
\usepackage{color}
\newcommand{\magsec}{{\rm mag/arcsec^2}}
\newcommand\dmpet{{\Delta m_{\rm p}}}
\newcommand\dmiso{{\Delta m_{\rm 25}}}
\newcommand\dmone{{\Delta m_{\rm 1\%}}}

\def\phiPB{{\phi_{\rm p, \,Blanton}}}

\begin{document}

\title{Photometric properties and luminosity function of nearby massive early-type galaxies}

\author{Y. Q. He\altaffilmark{1, 2, 3},
X. Y. Xia\altaffilmark{3},
C. N. Hao\altaffilmark{3},
Y. P. Jing\altaffilmark{4},
S. Mao\altaffilmark{2, 5},
Cheng Li\altaffilmark{6}
}

\altaffiltext{1}{University of Chinese Academy of Sciences, Beijing 100049, China}
\altaffiltext{2}{National Astronomical Observatories, Chinese Academy of Sciences, 20A Datun Road, Chaoyang District, Beijing 100012, China}
\altaffiltext{3}{Tianjin Astrophysics Center, Tianjin Normal University, Tianjin 300387, China; E-mail: xyxia@bao.ac.cn}
\altaffiltext{4}{Center for Astronomy and Astrophysics,
  Department of Physics and Astronomy, Shanghai Jiao Tong University, Shanghai
  200240, China}
\altaffiltext{5}{Jodrell Bank Centre for Astrophysics, University of Manchester, Alan Turing Building, Manchester M13 9PL, UK}
\altaffiltext{6}{Partner Group of the Max Planck Institute
  for Astrophysics at the Shanghai Astronomical Observatory and
  Key Laboratory for Research in Galaxies and Cosmology of Chinese
  Academy of Sciences, Nandan Road 80, Shanghai 200030,
  China}

\begin{abstract}

We perform photometric analyses for a bright early-type galaxy (ETG) sample
with 2949 galaxies ($M_{\rm r}<-22.5$ mag) in the redshift range of 0.05
to 0.15, drawn from the SDSS DR7 with morphological classification from Galaxy Zoo 1.
We measure the Petrosian and isophotal magnitudes, as well as the corresponding half-light
radius for each galaxy. We find that for brightest galaxies ($M_{\rm r}<-23$ mag), 
our Petrosian magnitudes, and isophotal magnitudes to 
25 ${\rm mag/arcsec^2}$ and 1\% of the sky brightness are on
average 0.16 mag, 0.20 mag, and 0.26 mag brighter than the SDSS Petrosian values, respectively.
In the first case the underestimations are caused by overestimations in the sky background 
by the SDSS PHOTO algorithm, while the latter two are also due to deeper photometry. 
Similarly, the typical half-light radii ($r_{50}$) measured by the SDSS algorithm are smaller than our measurements. 
As a result, the bright-end of the $r$-band luminosity function is found to decline more slowly than previous works. 
Our measured luminosity densities at the bright end are more
than one order of magnitude higher than those of Blanton et al. (2003), and
the stellar mass densities at $M_{\ast}\sim 5\times10^{11} M_{\odot}$ and $M_{\ast}\sim 10^{12} M_{\odot}$ are a few tenths 
and a factor of few higher than those of Bernardi et al. (2010). 
These results may significantly alleviate the tension in the assembly of massive galaxies 
between observations and predictions of the hierarchical structure formation model.

\end{abstract}

\section{INTRODUCTION}

The properties of nearby early type galaxies (ETGs), especially the
brightest ones, offer important clues to understanding the cosmic assembly history of
massive galaxies.  In terms of their morphologies, colors, stellar
population content and scaling relations,  ETGs appear to be relatively simple
systems compared with spirals and other galaxies. 
There are, however, a lot of renewed interests in their dynamical properties, particularly from the recent integral field unit surveys 
(e.g. Emsellem et al. 2007; Brough et al. 2010; Cappellari et al. 2012). There is also
 a debate on their stellar mass assembly processes (Renzini 2006; Scarlata et al. 2007), and so the understanding of ETGs remains a particularly interesting issue.

In the past two decades, the concept of ``downsizing'' (Cowie et al. 1996; Gavazzi \&
Scodeggio 1996; Fontanot et al. 2009) for galaxy formation has been widely discussed. According to this scenario, the
epoch of star formation in ETGs depends on the galaxy mass, namely, the most
massive ETGs formed their stars on a shorter time scale and 
at earlier times.  Furthermore, large ground-based and
space-based imaging and spectroscopic surveys at low and high redshifts
found that the mass function shows a weak evolution for massive ETGs since 
redshift $\sim 1$ (e.g. Cimatti et al. 2006; Scarlata et al. 2007; Cool et al. 2008; Cimatti 2009; Vulcani et al. 2011). 
In contrast, theory predicts that the typical stellar mass of most massive ETGs increases 
by a factor of 2-4 since redshift one (see De Lucia \& Blaizot 2007; Tonini et al. 2012). 

The size evolution for ETGs has also been the topic of many recent studies. It
has been shown that the massive and quiescent ETGs with stellar mass $M_{\ast}
\ge 10^{11}M_{\odot}$ at high redshift are more compact with effective radii a
factor of $\sim$ 3-5 smaller than the present-day ETGs with similar stellar
masses, but the stellar masses increase by a factor of $\sim$ 2 since redshift
2 (e.g. Daddi et al. 2005; Trujillo et al. 2006; van Dokkum et al. 2008).  A
popular explanation for the evolution of massive ETGs is minor dry mergers,
although additional physical mechanisms may be required (e.g. Cimatti et al.
2008; Bezanson et al. 2009; van Dokkum et al. 2010; Bruce et al.  2012 and
McLure et al. 2012). Moreoever, Valentinuzzi et al. (2010) and Poggianti
et al. (2013) found compact (superdense) massive ETGs in the local universe that have sizes
comparable to high-z massive galaxies. They argued that the evolution in
the median and average size is mild from high to low redshift for galaxies with stellar masses
$3\times 10^{10}M_{\odot} < M_{\ast} < 4\times 10^{11}M_{\odot}$, whereas the
evolution can still be substantial for ETGs more massive than $4\times 10^{11}M_{\odot}$.  
It should also be noted that such studies have not
focused on the evolution of ETGs at the high mass end with $M_{\ast} \ge
10^{12}M_{\odot}$.  As pointed out by Naab (2012), the most massive ETGs or
their progenitors start forming their stars at redshift $\sim 6$ or even earlier
and these high redshift galaxies are just the cores of their local
counterparts. The assembly at later times may be inside out by accretion of
stellar mass in the outskirts of these galaxies. However, the mass assembly
history of most massive ETGs is not yet completely clear: the current
$\Lambda$CDM model cannot reproduce the stellar age, metallicity and color
evolution simultaneously (Tonini et al. 2012; De Lucia \& Borgani 2012).  

To understand the mass and size evolution, accurate photometry for massive ETGs is an essential first step. Bright galaxies tend to be located in crowded environments or have extended stellar halos. Inadequate masking of their neighbors or extended halos can result in an overestimate of the 
sky background, leading to an underestimate of their luminosity and size.
Indeed a number of investigators have noted that the SDSS photometric
reduction systematically underestimates the luminosities and half-light radii
for bright ETGs (see Aihara et al. 2011 for the SDSS DR8 release and
DR7 documentation). For the SDSS DR8, 
a more sophisticated sky background subtraction algorithm has been adopted 
and significant improvements were achieved, but some problems still persist 
(see \S4.1). On the other hand, the surface brightness distribution of 
most massive ETGs
cannot be fit well by a simple model such as the commonly used S\' ersic (1963) law, which can also lead to an underestimate of the luminosities.
Graham et al. (2005) carefully investigated the deficiency of
Petrosian (1976) system to quantify galaxy luminosity and size. They find that the Petrosian magnitude and size commonly used by studies based on SDSS strongly depend on the surface brightness profile.  For example, the flux deficiency can be as large as 0.5 mag for a galaxy
with $r^{1/8}$ surface brightness profile. To avoid underestimating the
luminosities of the brightest cluster galaxies (BCGs), von der Linden et al.
(2007) used isophotal photometry to 23 ${\rm mag/arcsec^2}$ in the $r$-band, rather
than the Petrosian magnitude. Therefore, aperture photometry offers a
viable alternative to describe the photometric properties of ETGs.  

Underestimating the luminosities of luminous ETGs will lead to an
underestimate of the stellar mass density at the bright end. Using their
own sky background subtraction algorithm (cmodel) and  stellar mass estimation,
Bernardi et al. (2010) found a higher number of very massive galaxies than 
previous works. The excess can be up to a factor of 
$\sim$ 10 when the stellar mass $M_{\ast}$ is larger than $5\times 10^{11}M_{\odot}$, which is highly significant.

In this work, we perform accurate photometry for a complete
sample of nearby bright early-type galaxies with $M_{\rm r}<-22.5$ mag and
$M_\ast > 10^{11}M_{\odot}$. The morphologies of these galaxies are taken from the Galaxy Zoo project which classified nearly
900,000 SDSS galaxies (Lintott et al. 2011). We perform our own sky background subtraction, which leads to much more accurate photometry of ETGs. We then compare the
luminosities and sizes measured by different methods and investigate the
luminosity function as well as the stellar mass density for the bright ETGs.

The structure of this paper is as follows. In \S2 and \S3 we describe the sample
selection and data reduction for our bright ETGs. In \S4, we
compare our measured luminosities and sizes of bright ETGs with the SDSS Petrosian
magnitudes and sizes, and present the results of the $r$-band luminosity function
and the stellar mass density. We finish the paper with a summary in \S5. In this
paper, we adopt a Hubble constant of $H_{\rm 0}=70\,{\rm km
\, s^{-1} Mpc^{-1}}$, matter density parameter $\Omega_{m}=0.3$ and
cosmological constant $\Omega_{\Lambda}=0.7$.

\section{SAMPLE}

Our bright early type galaxy (ETG) sample is drawn from the morphological
catalogue of Galaxy Zoo 1 (Lintott et al. 2011). The Galaxy Zoo 1 project
performed visual morphological classifications for nearly 900,000 galaxies
from the Sloan Digital Sky Survey (SDSS, York et al. 2000), in which 667,945
galaxies are from the Main Galaxy Sample of SDSS (MGS, Strauss et al. 2002). The MGS
includes galaxies with $r$-band Petrosian magnitude $r\le17.77$ and $r$-band
Petrosian half-light surface brightness $\mu_{50} \le 24.5$ {\rm
mag/arcsec}$^2$. All 667,945 galaxies from the MGS have spectroscopic redshift
in the range of $0.001<z<0.25$ and u, g, r, i, z band photometry based on SDSS
DR7 (Lintott et al. 2011).

Given that the SDSS spectroscopy survey is incomplete for bright galaxies with
redshift less than 0.05 (Stoughton et al. 2002; Strauss et al. 2002; Schawinski
et al. 2007; Kaviraj et al. 2007) and reliable photometric analyses with
high signal to noise (S/N) ratios can only be performed for galaxies 
with $r\le16$ mag (Fukugita et al. 2007), we constrain our bright ETG
sample to $z>0.05$ and $r<16$ mag.  Since
we are only concerned with the photometric properties of the luminous ETGs, we
further restrict our sample to bright ETGs with $M_{\rm r}<-22.5$ mag. The Petrosian absolute magnitudes, $M_{\rm r}$, are 
calculated using the equation $M_{\rm r}=m_{\rm r}-5 \log(D_{\rm L}/10\,{\rm pc})-A-k$,
 where $D_{\rm L}$ is the luminosity distance, 
$A$ is the Galactic extinction obtained from the photometric
catalogue of SDSS DR7 and $k$ is the k-correction derived using the IDL KCORRECT
algorithm of Blanton et al. (2007).  Finally there are 7930 bright ETGs in the
redshift range of $0.05<z<0.15$ and $M_{\rm r}<-22.5$ mag.

\begin{figure}
\includegraphics[width=1.0\textwidth]{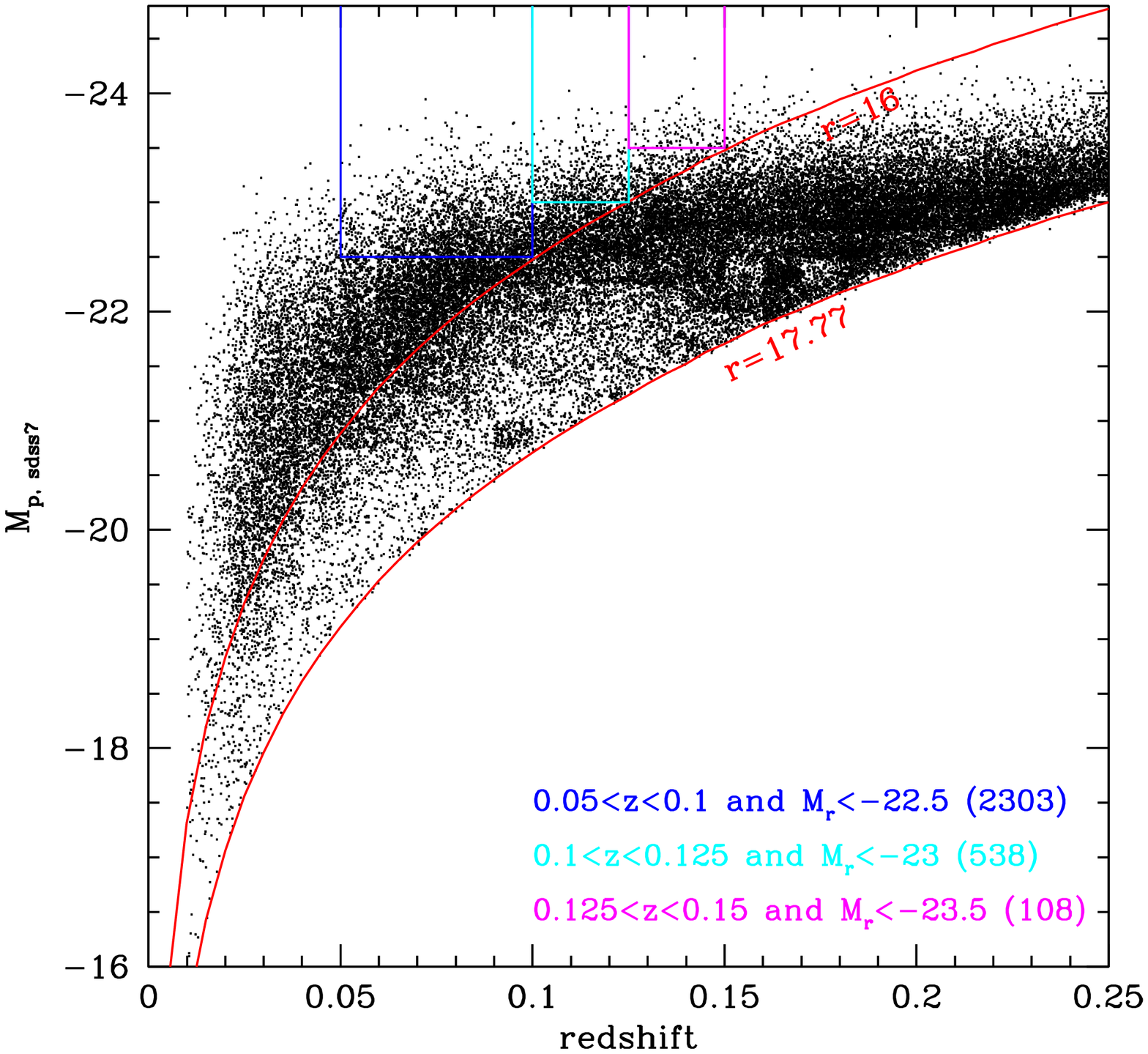}
\caption{Three volume-limited subsamples selected from Galaxy Zoo 1
  early-type galaxies in the redshift vs. absolute Petrosian magnitude
  plane in the $r$-band of SDSS DR7, 
shown in blue, cyan and magenta boxes, respectively. 
The total number of galaxies is 2949 ETGs. 
The red lines show the observed flux limits at $r=16$ mag and 17.77 mag, respectively.
The number of early-type galaxies in each subsample within the redshift range and 
lower luminosity limit are shown in the bottom right.
\label{sample-fig1}}
\end{figure}

\begin{figure}
\centering
\includegraphics[width=1.0\textwidth]{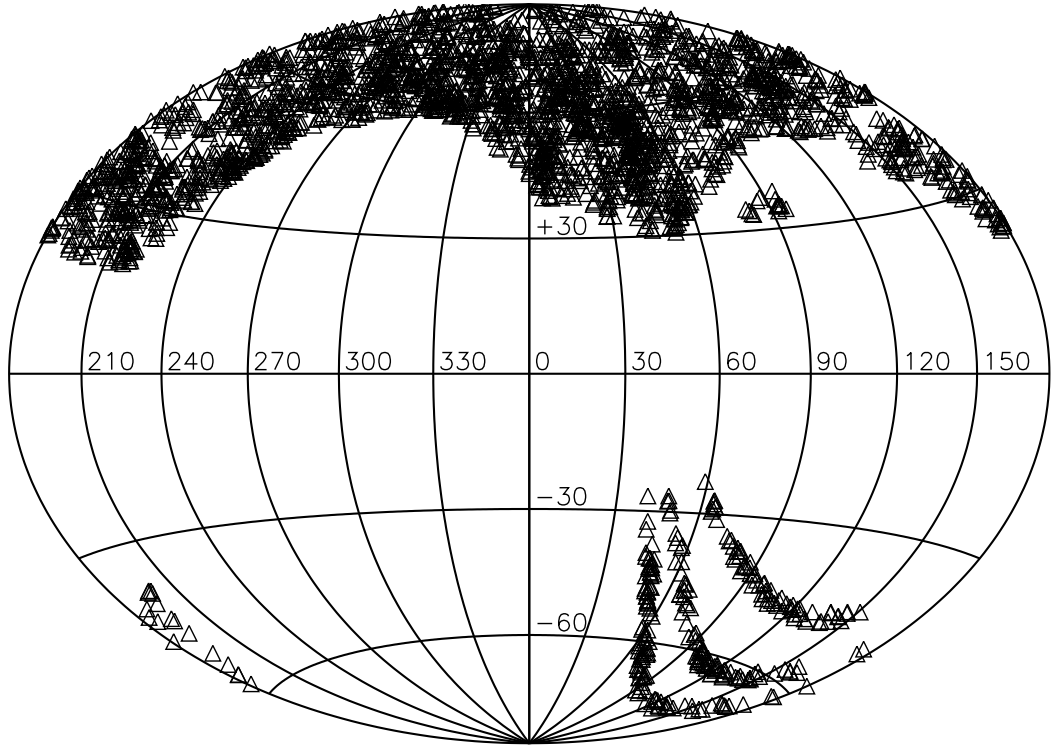}
\caption{The sky coverage for our volume-limited bright early-type galaxies in the Aitoff projection in Galactic coordinates.
The total sky coverage is about 9055 degree$^2$. 
}
\label{sky-fig2}
\end{figure}

\begin{figure}
\includegraphics[width=1.0\textwidth]{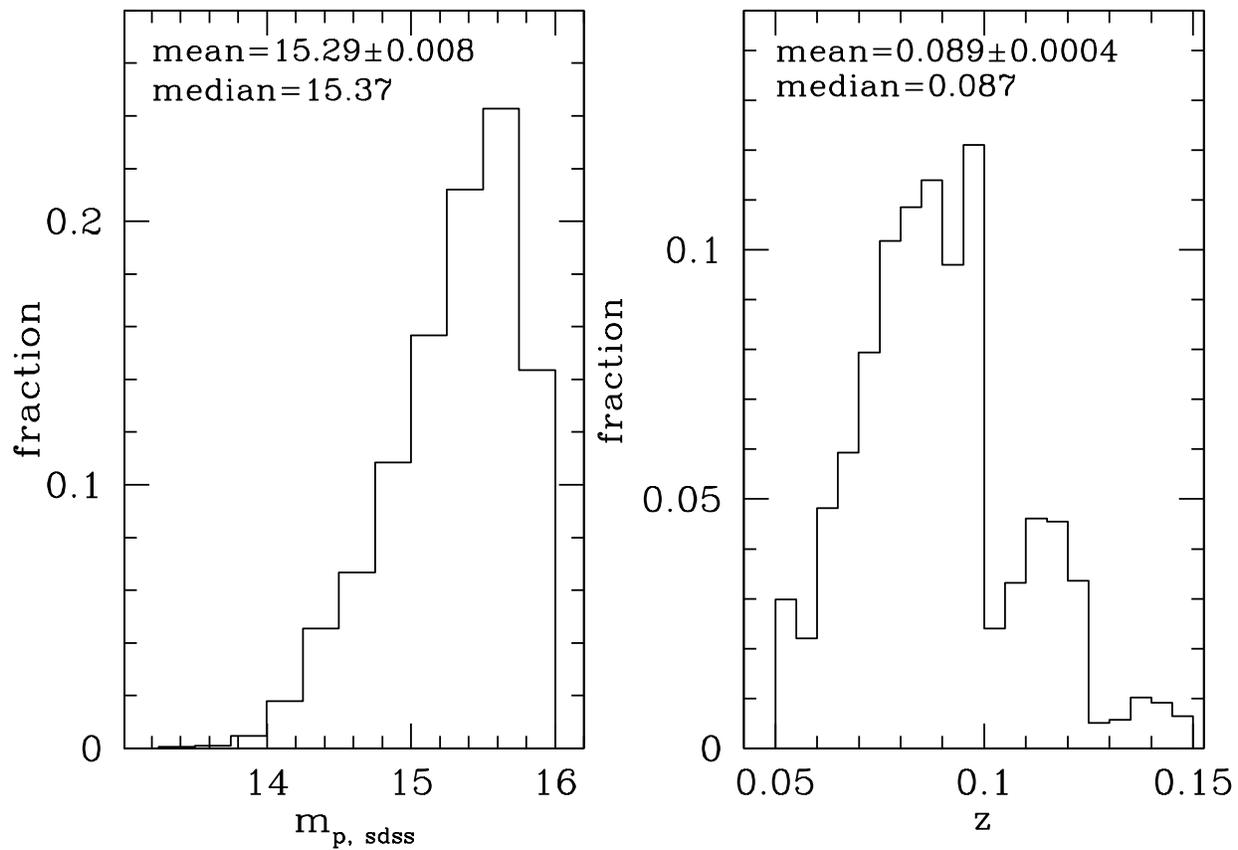}
\caption{Distributions of the SDSS Petrosian apparent magnitude in the $r$-band (left panel) and redshift
(right panel) for 2949 early-type galaxies with $M_{\rm r}<-22.5$ mag. 
The mean and median values for each distribution are shown in the top left.
\label{histzmag-fig3}}
\end{figure}

We further divide this sample into three volume-limited subsamples 
with 2303 ETGs in the redshift ranges of $0.05<z<0.1$ for $M_{\rm r}<-22.5$ 
mag, 538 ETGs with $0.1<z<0.125$ for $M_{\rm r}<-23$ mag
 and 108 ETGs with $0.125<z<0.15$ for $M_{\rm r}<-23.5$ mag, as
shown in Fig.~\ref{sample-fig1}.  Then we perform $<V/V_{\rm max}>$ test (Schmidt
1968) for these subsamples.  The average value $<V/V_{\rm max}>$ are 0.51,
0.50 and 0.51 for the three subsamples, respectively, which indicates that all 
the three subsamples are spatially homogeneous.
The photometric analysis in this work is based on these three subsamples. 
In total, there are 2949 ETGs brighter than $-$22.5 mag, among which 1053 ETGs
are brighter than $-$23 mag. 
Fig.~\ref{sky-fig2} shows the spatial distribution of our sample
galaxies on the sky in Galactic coordinates: the total area is 9055 degree$^2$,
corresponding to 22\% of the whole sky.  Fig.~\ref{histzmag-fig3} presents the
histograms showing the distributions of apparent magnitude and spectroscopic
redshift for 2949 early-type galaxies. The median Petrosian magnitude is 15.37
mag and median redshift is 0.087, respectively.

In order to verify the early-type morphology of our sample galaxies, we
visually inspected the SDSS $r$-band images for all 2949 galaxies. In
addition, we also checked other commonly used classification  criteria for
early-type galaxies.  All our sample galaxies have $g-r>0.7$
(e.g. Blanton et al. 2003; Shen et al. 2003),  and most (94\%) have 
concentration index $C_{r}=r_{90}/r_{50}$ larger than 2.86 (Bernardi et al. 2010), where $r_{90}$ and $r_{50}$ are the radii containing 90\% and 50\% of Petrosian flux. 
Therefore, our selected sample galaxies are all ETGs.

\section{DATA REDUCTION}

\subsection{Estimation of the Sky Background} \label{}

The corrected frame fpC-images in the $r$-band for our sample galaxies are directly
obtained from the SDSS DR7 Data Archive Server. The images have been
pre-processed by the SDSS photometric pipeline (PHOTO), which includes bias
subtraction, flat-fielding and bad pixels correction (cosmic rays removal, bad
columns and bleed trails).  In order to obtain the SDSS photometric flux
calibration information during observations, such as the photometric zeropoint
$a$, the first-order extinction coefficient $k$ and the airmass $X$, we also
downloaded the calibrated field statistic file, named as tsField from the archive.

As mentioned in \S1, the SDSS photometric reduction systematically
underestimates the luminosities and half-light radii for bright ETGs, 
which is mainly caused by inadequate masking of their neighbors or extended 
stellar halos (Aihara et al. 2011), leading to an overestimate of 
the sky background. 
In this work we perform the masking more carefully and estimate the sky background 
model following Liu et al. (2008), which has been
successfully used to measure the luminosities and half-light radii for the
brightest cluster galaxies (BCGs) that are located in crowded fields and
usually have extended faint stellar halos. 

\begin{figure}
\centering
\includegraphics[width=0.8\textwidth]{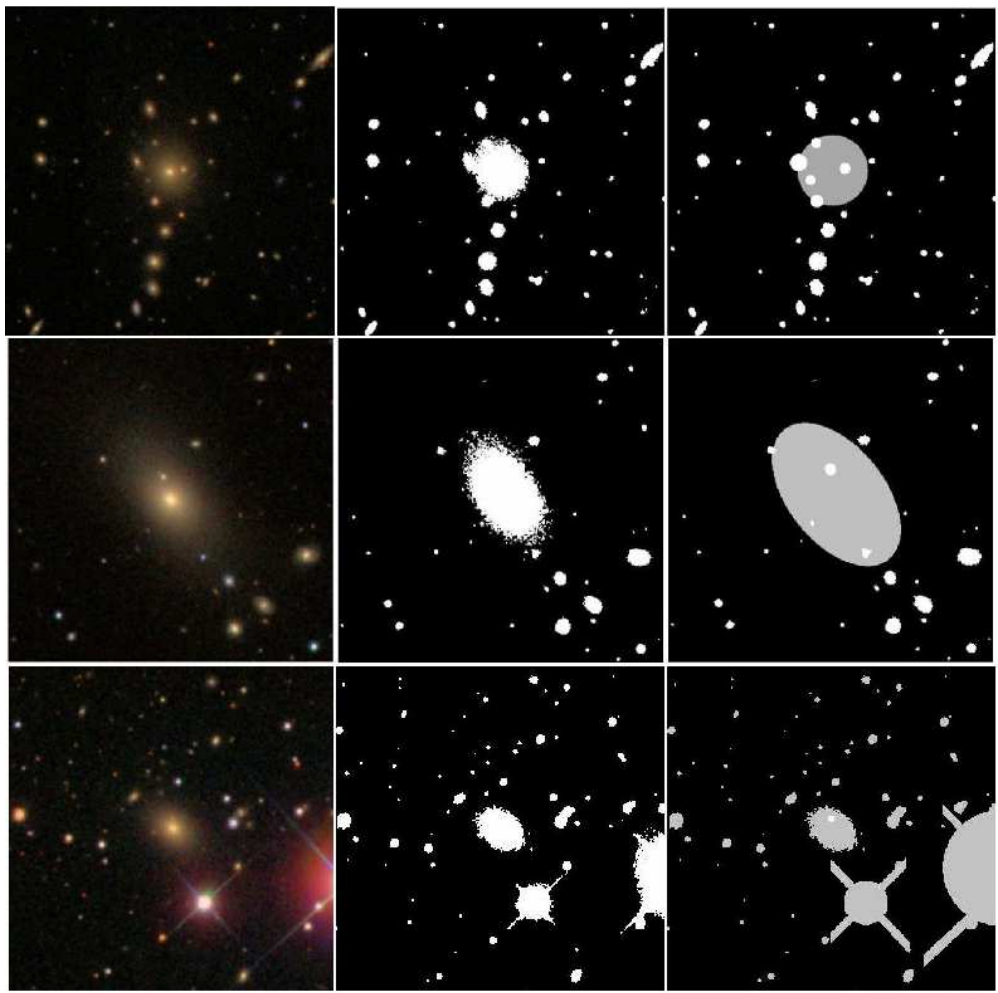}
\caption{Examples of objects that cannot be masked well by the automatic algorithm of SExtractor.
The left panels show examples of color images of ETGs in a crowded field (top row),
with extended stellar halo (middle row), and with nearby contaminated stars (bottom row).
The middle and right columns show the corresponding masked images generated by 
SExtractor and by hand. The white areas indicate masked regions.
}
\label{handmask-fig4}
\end{figure}

In the following, we outline our sky background subtraction approach. 
In order to obtain the sky background, we first masked out all objects detected 
by SExtractor (Bertin \& Arounts 1996) in the corrected frames with $2048 \times 1489$ pixels ($13\arcmin.5 \times
9\arcmin.8$). We then carefully checked each frame by eye to make sure that
the wings of bright stars or the faint stellar halos of galaxies have been properly
masked. As expected, we find that the automatic algorithm of
SExtractor does not work well for about 24\% ETGs brighter than $-$22.5 mag
and 33\% ETGs brighter than $-$23 mag, which mostly include ETGs residing in crowded fields, 
having extended stellar halos or are close to foreground bright stars. For the 2949 ETGs
brighter than $-$22.5 mag, the percentages of these objects are about 13\%,
7\%, 4\%, respectively. For the 1053 ETGs brighter than $-$23 mag, the
percentages are even higher, about 18\%, 10\%, 5\%, respectively. For these ETGs, we modify the mask images manually.  
For comparison, Fig.~\ref{handmask-fig4} shows three cases (top, middle and bottom rows) between the
masked images generated by the automatic algorithm of SExtractor (middle
panels) and those by hand (right panels). The left panels of Fig.~\ref{handmask-fig4} show the original
true color images for the example galaxies. The top row shows a galaxy in
a crowded field.  From the top middle panel, it can be seen that SExtractor
cannot separate the target galaxy from its neighbors 
well. Therefore we manually flag the nearby objects with circles that are large
enough to cover the objects completely, as shown in the top right panel.  The
second row shows a bright ETG with an
extended stellar halo.  From the middle panel, it is clear that
the flagged area generated by SExtractor is not big enough to cover the whole
stellar halo. Hence, the stellar halo would be considered as the
background and subtracted from the galaxy itself, leading to underestimates
of the luminosity and half-light radius of the galaxy.  Following the
shape of the target ETG, we use an ellipse to mask the entire stellar envelope,
as shown in the middle right panel. SExtractor also cannot handle well 
galaxies surrounded by bright foreground stars, especially those
with diffractive spikes (bottom panels).  Therefore, we use long
rectangles to mask the star spikes (bottom right). The
masked images generated manually can provide not only  more accurate sky
background images but also good masks for the surface photometry (see \S 4).

To increase the valid area of the sky background subtraction, we smooth the sky background only image with a median filter of $51
\times 51$ pixels. 
This filter size is selected to be larger than the sizes of most objects in the frame but still sufficiently small so that the variation of the sky background within the region is still reproduced (i.e., not smoothed out).
After the median filtering is performed, masked regions
smaller than the median box filter are replaced with the surrounding
sky background, whereas part of the masked regions larger than the median box
filter remain flagged.  
With the small field of view ($13\arcmin.5 \times 9\arcmin.8$), the
  filtered sky background image is fitted with a two-dimensional first-order 
Legendre polynomial (i.e., $z=a+b*x+c*y$) using the IRAF/IMSURFIT task.
The sky background map is typically tilted with a spatial variation of $\sim 1-2$ ADU across
the whole frame (Liu et al. 2008).  We subtract this
sky background model from the initial fpC-image to obtain the sky-corrected frame.
The blank regions in the sky-subtracted frame follow Gaussian distributions
with means close to zero and standard deviations of several ADU, which
is consistent with Liu et al. (2008).  After the sky subtraction, we trim both the
sky-subtracted frame and the mask image to $501 \times 501$ pixels centered on the
galaxy of interests. The trimmed mask image with the target galaxy
un-flagged will be used to probe the masked regions in the isophote fitting.

\subsection{Isophotal Photometry}

After the sky background subtraction, we perform surface photometry for our
bright ETGs in order to estimate the luminosities and sizes of sample galaxies.
As is well known, the Petrosian magnitude and Petrosian size are most commonly
used to describe the galaxy flux and half-light radius for the analyses based
on SDSS database, because they do not depend on the model fitting to galaxies.
The Petrosian radius $r_{\rm p}$ is defined as the radius $r$ at which the ratio of
the local surface brightness averaged over an annulus between $0.8r$ and
$1.25r$ to the mean surface brightness within $r$ equals to 0.2 and the
Petrosian magnitude is the integrated flux within $2r_{\rm p}$ (Petrosian 1976).
However, the Petrosian magnitude misses the light outside 
$2r_{\rm p}$ (Petrosian
aperture) because of its dependence on the surface brightness profile of
galaxies (Graham et al. 2005), which leads to underestimates of the fluxes and sizes
for galaxies with extended stellar halos.  In this work, we
will measure not only the Petrosian magnitude and Petrosian half-light radius,
but also the isophotal magnitude and half-light radius to deeper
isophote limits at 25 ${\rm mag/arcsec^2}$ and 1\% of sky brightness, 
corresponding to $\sim$ 26 ${\rm mag/arcsec^2}$ in most cases (Bernardi et al.
2007). 

We perform the surface photometry analysis following Wu et al. (2005) and Liu
et al. (2008). The procedures are briefly described below. First, we use {\tt
ISOPHOTE/ELLIPSE} task in IRAF to fit each of the trimmed sky-subtracted images
excluding the masked regions in the fitting. The surface brightness of
the target galaxy is fitted by a series of elliptical annuli in a logarithmic
step of 0.1 along the semi-major axis. The annuli chosen in the outer
parts of image is larger, which can suppress the shot noise in the outer
regions where the signal to noise ratio (S/N) is much lower.  The output of
{\tt ISOPHOTE/ELLIPSE} is the mean intensity in each isophote annulus.  Then we
integrate the surface brightness profile to isophotal limits of 25 ${\rm
mag/arcsec^2}$ or 1\% of sky brightness to obtain the apparent
magnitudes and half-light radius $r_{50}$.  We also measure the Petrosian
magnitudes and $r_{50}$ based on our sky background subtracted
images.  In our analysis, the equivalent radius $\sqrt {ab}$ of an ellipse is
used for all the radial profiles, where $a$ and $b$ are the semi-major and
semi-minor radii of the ellipse. The cosmological dimming is also taken into
account for the surface brightness profiles.  The observational errors in the
surface brightness profile include random errors (e.g., the shot noise
of the object and sky background, readout noise and noise
contributed by data reduction) and the error from sky background subtraction.

\section{RESULTS}

\subsection{Luminosities of bright ETGs in the local universe}

The SDSS database provides the largest galaxy sample with both photometric and
spectroscopic information in the local universe. However, the underestimation
of the luminosities and sizes for the bright ETGs prevents us from
correctly understanding their properties and assembly history. 

For brevity in later discussions, we define
\begin{equation}
\begin{array}{l}
\displaystyle \dmpet=m_{\rm p, \,sdss7}-m_{\rm p}, \\
\displaystyle \dmiso=m_{\rm p, \,sdss7}-m_{25},\\
\displaystyle \dmone=m_{\rm p, \,sdss7}-m_{1\%},
\label{eq:dm}
\end{array} 
\end{equation}
where $m_{\rm p}$, $m_{25}$ and $m_{1\%}$ are our measured Petrosian magnitude, 
isophotal magnitudes with surface brightness measured to 25 mag/arcsec$^2$ and 1\% of sky brightness, 
and $m_{\rm p, \,sdss7}$ is the Petrosian magnitude from the SDSS DR7 pipeline in the $r$-band.

\begin{figure}
\centering
\includegraphics[width=0.78\textwidth]{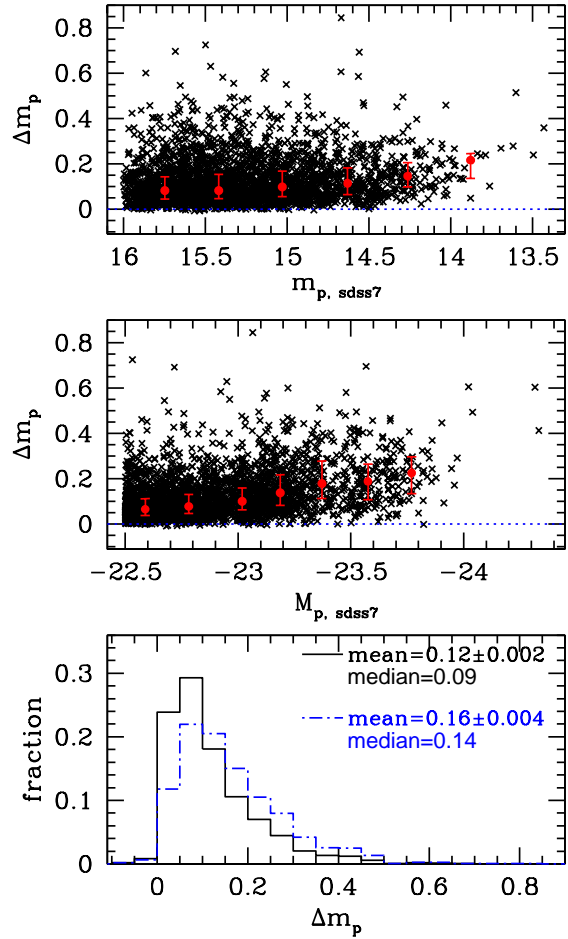}
\caption{The difference $\dmpet$ between the SDSS Petrosian magnitude and our measured
Petrosian magnitude as a function of the SDSS apparent Petrosian magnitude $m_{\rm p, \,sdss7}$
(top panel) and Petrosian absolute magnitude $M_{\rm p, \,sdss7}$ (middle panel).
The red data points with error bars
are the median, lower (25 per cent) and upper (75 per cent) quartiles for binned 
galaxies. The bin width is 0.4 mag for the apparent Petrosian magnitude except the
last bin which has a width of 0.6 mag to include all the remaining objects (top panel);
the bin width is 0.2 mag for the absolute magnitude except the last bin which has a
width of 0.7 mag to include all the remaining objects (middle panel).
The medians shown in red points are 0.081, 0.081, 0.098, 0.114, 0.145, 0.216 mag in top panel,
and 0.065, 0.078, 0.101, 0.138, 0.179, 0.188, 0.225 mag in middle panel.
The bottom panel shows the histograms of the magnitude difference with
the mean and median values indicated in the top right, 
as black solid line for galaxies with $M_{\rm p, \,sdss7}<-$22.5 mag and blue dot dashed line for galaxies with $M_{\rm p, \,sdss7}<-$23 mag.
}
\label{1053mpetrodiff-fig5}
\end{figure}

The top and middle panels in Fig.~\ref{1053mpetrodiff-fig5} show
$\dmpet$ (defined eq. \ref{eq:dm}) as a function of the SDSS Petrosian apparent 
and absolute magnitudes, respectively, while the bottom panel shows 
the histogram of $\dmpet$. Clearly for more luminous ETGs, 
the luminosity difference between the SDSS and our measurements is larger. 
The mean and median values of the luminosity differences are 0.16 mag and 
0.14 mag, respectively. It shows that the algorithm used in the SDSS DR7 has systematically
overestimated the sky background for bright ETGs, leading to underestimates in
the luminosities of bright ETGs, especially for the brightest ones.

\begin{figure}
\centering
\includegraphics[width=1.0\textwidth]{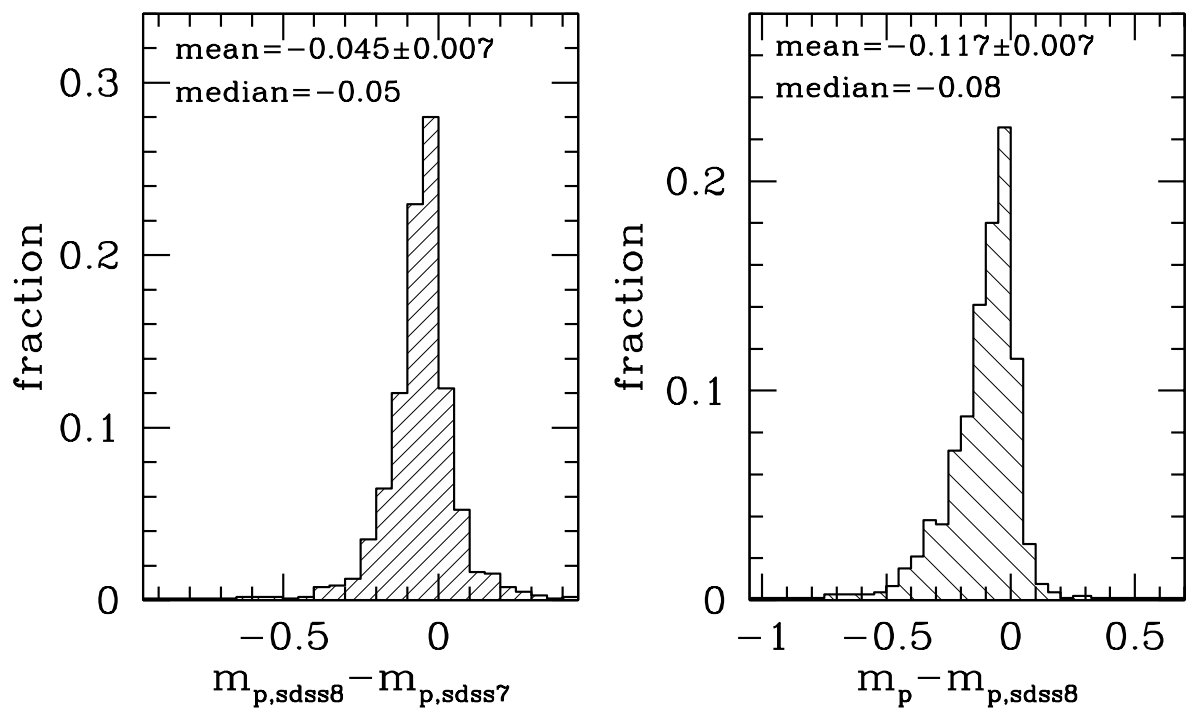}
\caption{The distributions of the difference between the SDSS DR8 and SDSS DR7 
Petrosian magnitudes (left panel), and the
difference between our measured Petrosian magnitude and the SDSS DR8 (right panel).
The mean and median values for each distribution are shown in the top left of the panel.}
\label{comparewithdr8-fig6}
\end{figure}

Addressing the same issue, in the SDSS DR8, Aihara et. al. (2011) have re-processed
all SDSS imaging data using a more sophisticated sky background subtraction
algorithm and obtained significant improvement. The left and right panels
of Fig.~\ref{comparewithdr8-fig6} are histograms of luminosity difference
 between the SDSS DR8 and SDSS DR7, and between our
measurements and SDSS DR8, respectively.  We can see 
that the median values of luminosity difference for these two cases are $-$0.05 mag and $-$0.08 mag, and the mean values are about $-$0.04 mag and $-$0.12 mag, respectively.
The new algorithm on sky background subtraction used in the SDSS
DR8 has indeed improved the photometric measurement (the mean shift is about 0.04 mag). However, compared to our study it still
underestimates the luminosities by about 0.12 mag on average for bright ETGs.
We carefully checked each image with very large luminosity
differences between the SDSS DR8 and our measurements. We find that most of these
galaxies reside in crowded fields or have very extended faint stellar
halos, and so the SDSS DR8 treatment of the sky background issue is incomplete.

\begin{figure}
\centering
\includegraphics[width=0.9\textwidth]{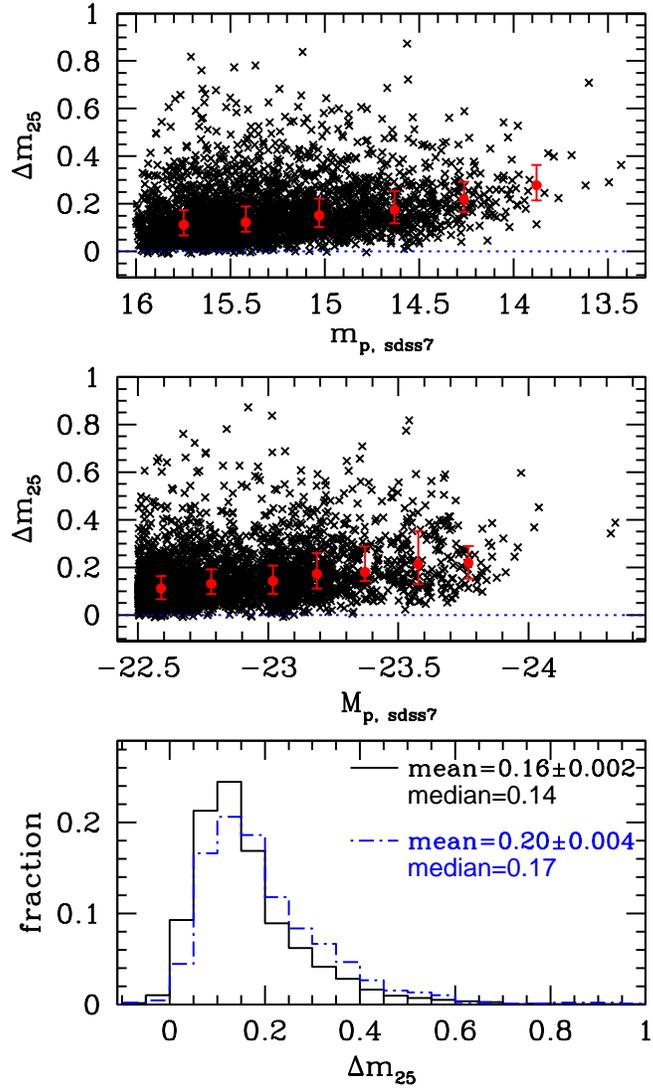}
\caption{The difference $\dmiso$ between the SDSS Petrosian magnitude and our 
isophotal magnitude to 25 mag/arcsec$^2$ as a function of the SDSS apparent 
$m_{\rm p, \,sdss7}$ (top panel) and absolute $M_{\rm p, \,sdss7}$  (middle panel) isophotal magnitudes.
See Fig.~\ref{1053mpetrodiff-fig5} for an explanation of the red data points
with error bars in the top and middle panels.
The medians shown in red points are 0.112, 0.122, 0.151, 0.175, 0.220, 0.278 mag in top panel, and 0.111, 0.131, 0.142, 0.172, 0.180, 0.213, 0.219 mag in middle panel.
The bottom panel shows the histogram of magnitude difference distribution with
the mean and median values indicated in the top right,
as black solid line for galaxies with $M_{\rm p, \,sdss7}<-$22.5 mag and blue dot dashed line for galaxies with $M_{\rm p, \,sdss7}<-$23 mag.
}
\label{1053misosdssdiff-fig7}
\end{figure}

Liu et al. (2008) found that the isophotal magnitudes $m_{25}$ measured
to the surface brightness of 25 ${\rm mag/arcsec^2}$ are generally larger than
the Petrosian values for BCGs. We thus also compare the Petrosian and isophotal
magnitudes for our sample ETGs. The top and middle panels of
Fig.~\ref{1053misosdssdiff-fig7} show $\dmiso$ (defined in eq. \ref{eq:dm}), as a function of
the SDSS Petrosian apparent and absolute magnitudes, respectively. It is obvious that
$\dmiso$ increases with the apparent and absolute magnitudes of
ETGs. For $M_{\rm p, \,sdss7}<-23$ mag, the mean and median differences are 0.20 mag and 0.17 mag,
respectively, which are larger than those for $\dmpet$ (0.16 and 0.14 mag, respectively). 

\begin{figure}
\centering
\includegraphics[width=0.9\textwidth]{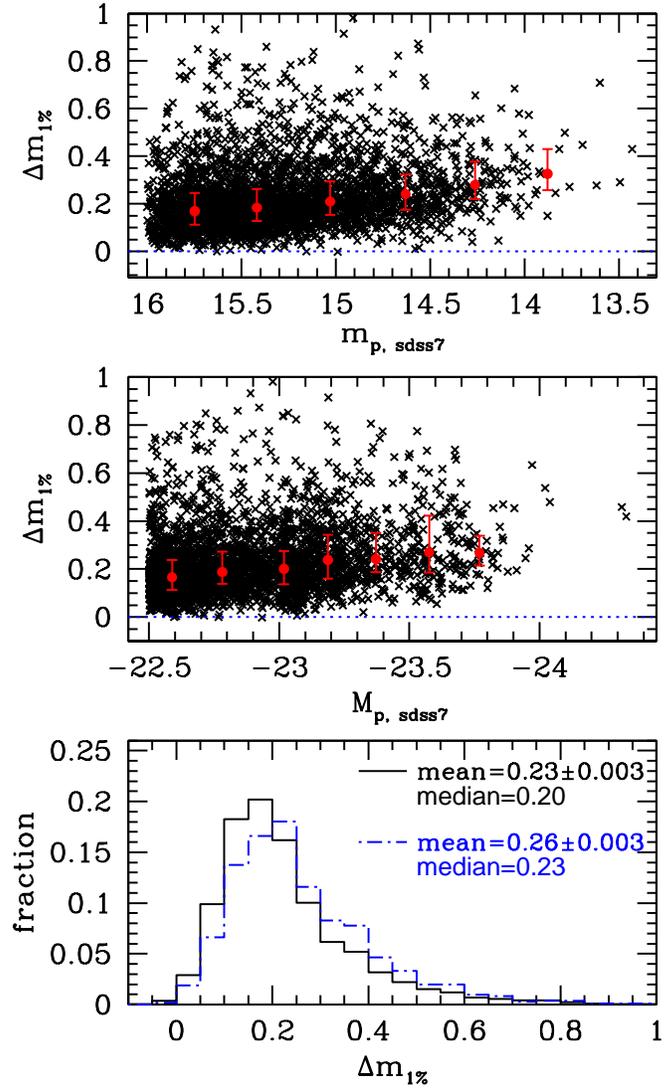}
\caption{The difference $\dmone$ between the SDSS Petrosian magnitude
and our isophotal magnitude with surface brightness limit at 1\% of sky brightness
as a function of the SDSS apparent $m_{\rm p, \,sdss7}$ (top panel) and 
absolute $M_{\rm p, \,sdss7}$ (middle panel) magnitudes.
 See Fig.~\ref{1053mpetrodiff-fig5} for an explanation of the red data points
with error bars in the top and middle panels.
The medians shown in red points are 0.170, 0.184, 0.209, 0.241, 0.280, 0.326 mag in top panel, and 0.166, 0.131, 0.188, 0.201, 0.237, 0.242, 0.270, 0.267 mag in middle panel.
The bottom panel shows the histogram of magnitude difference distribution with
the mean and median values indicated in the top right,
as black solid line for galaxies with $M_{\rm p, \,sdss7}<-$22.5 mag and blue dot dashed line for galaxies with $M_{\rm p, \,sdss7}<-$23 mag.
}
\label{1053mskysdssdiff-fig8}
\end{figure}

Bernardi et al. (2007) investigated the galaxy luminosity by integrating the
best-fit model of galaxy surface brightness profile to 1\% of sky
brightness.  The top and middle panels of Fig.~\ref{1053mskysdssdiff-fig8} show $\dmone$ 
(defined in eq. \ref{eq:dm}) as a function of
the SDSS Petrosian apparent and absolute magnitudes, respectively. 
The same trend is seen: $\dmone$ increases as ETGs become more luminous.
For $M_{\rm p, \,sdss7}<-23$ mag, the mean and median values of $\dmone$ are 0.26 mag and 0.23 mag, respectively.  
Note that the photometric limit of 1\% sky is $\sim$ 26 $\magsec$, almost one magnitude deeper than 25 $\magsec$.  

We further examine the percentage of ETGs with large
luminosity difference between the SDSS Petrosian and our measured magnitudes. For the 2949 ETGs brighter than $-22.5$
mag, there are 6\%, 12\% and 22\%  ETGs with $\dmpet$,
$\dmiso$ and $\dmone$ larger than 0.3 mag. For the 1053
ETGs brighter than $-23$ mag, the fractions are even higher, 11\%, 19\% and 31\%, respectively.
For more luminous ETGs, the underestimation
by the SDSS algorithm is more severe. In addition, the deeper the photometric measurements, the larger the underestimations.

We point out that to the photometric limit of $\sim$ 26 ${\rm mag/arcsec^2}$
the light still belongs to galaxies as shown
by Tal \& van Dokkum (2011). They stacked more than 42000
SDSS images of LRGs (Luminous Red Galaxies), reaching a depth of $\sim$30
${\rm mag/arcsec^2}$, and found that the stellar light out to 100
kpc is physically associated with galaxies, instead of inter-cluster 
or inter-group light. Our photometric measurement to 1\% sky brightness
reaches at most 100 kpc (50 kpc on average, see
\S4.2), and so our measured isophotal light is from ETGs themselves.

\subsection{Sizes of bright ETGs in the local universe}

Galaxy size is one of the most important parameters of galaxy
properties. A reliable determination of the sizes of bright ETGs in the local universe 
provides the basic calibration for investigating the size evolution, an area of active study 
in recent years (e.g. Szomoru et al. 2012; McLure et al. 2012;
Trujillo 2012). The aforementioned works compare galaxy sizes at high redshift to those of
the local universe by Shen et al. (2003) based on the SDSS PHOTO algorithm,
who found a power-law relation between the galaxy luminosity and size
(Shen et al. 2003). However, if the luminosities of galaxies have been
underestimated, galaxy sizes may have been underestimated too. In this section,
we will discuss the size difference between the SDSS and our measurements 
based on different luminosity estimations. For convenience, we define
$r_{\rm 50, \,p}$, $r_{\rm 50, \,25}$ and $r_{\rm 50, \,1\%}$ as
our Petrosian half-light radius, isophotal half-light radii to 25 mag/arcsec$^2$ and 1\% of 
the sky brightness. These values will be compared with $r_{\rm 50, \,sdss7}$, the Petrosian 
half-light radius from the SDSS DR7 pipeline in the $r$-band.
 
\begin{figure}
\centering  
\includegraphics[width=1.0\textwidth]{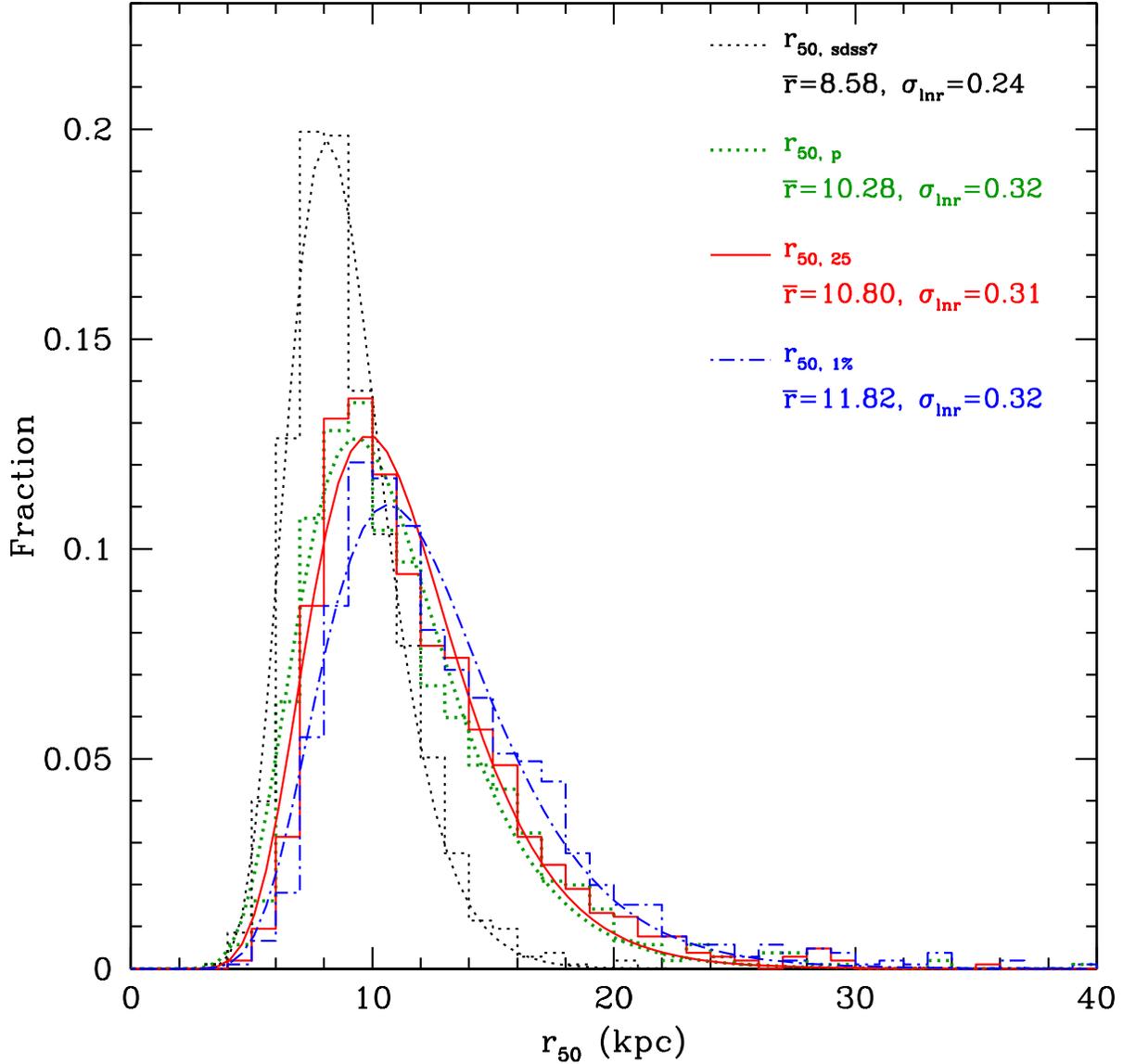}
\caption{The histograms of half-light radii ($r_{50}$) measured by different photometric
methods. The black and green dotted lines represent the Petrosian $r_{50}$ distributions 
based on SDSS and our own measurements, respectively.
While the red solid and blue dot-dashed lines show the $r_{50}$ distributions 
based on isophotal measurement to 25 ${\rm mag/arcsec^{2}}$ and 1\% of sky brightness, respectively. 
Log-normal (see eq. \ref{eq:lognormal}) fits are shown for each distribution with 
corresponding colors, and the medians and dispersions are shown in the top right.
}
\label{1053hist_r50-fig9}
\end{figure}

Fig.~\ref{1053hist_r50-fig9} presents histograms of the half-light
radii measured by different photometric methods. 
Following Shen et al. (2003), a log-normal function is fitted
to the half-light radii distribution. The log-normal function is defined as
\begin{equation}
f(r,\bar{r},\sigma_{\ln{r}})=\frac{1}{\sqrt{2\pi}\sigma_{\ln{r}}}
  \exp\left[-\frac{\ln^2(r/\bar{r})}{2\sigma_{\ln{r}}^2}\right]
  \frac{dr}{r},
\label{eq:lognormal}
\end{equation}
which is characterized by the median $\bar{r}$ and the dispersion $\sigma_{\ln{r}}$.
We find that the best-fit values for $r_{\rm 50, \,sdss7}$,
$r_{\rm 50, \,p}$, $r_{\rm 50, \,25}$, $r_{\rm 50, \,1\%}$ distributions are 
$\bar{r}=8.58, 10.28, 10.80, 11.82$ kpc and $\sigma_{\ln{r}}=0.24, 0.32, 0.31, 0.32$.
From Fig.~\ref{1053hist_r50-fig9}, it can be clearly seen that the largest $r_{\rm 50, \,sdss7}$ value
is smaller than 20 kpc, while our measured $r_{\rm 50, \,p}$,
$r_{\rm 50, \,25}$ and $r_{\rm 50, \,1\%}$ can be as large as 30 kpc and 
a large fraction ($\sim$ 27\%) of brightest ETGs have sizes larger than 15 kpc.  
Therefore, the SDSS algorithm has significantly underestimated the real sizes of  bright ETGs.

\begin{figure}
\centering
\includegraphics[width=0.92\textwidth]{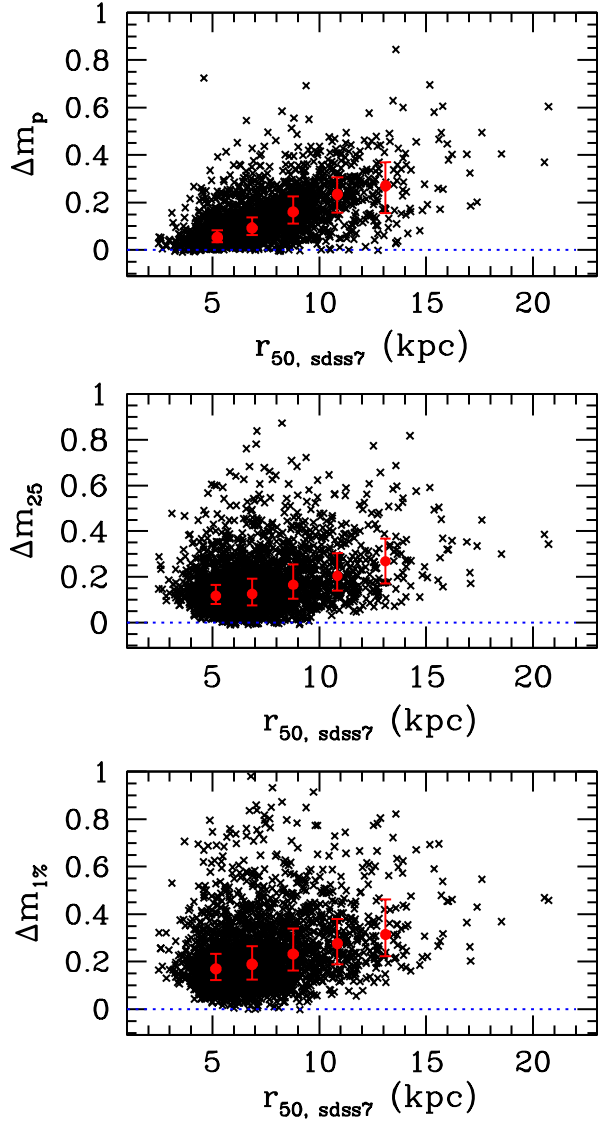}
\caption{The magnitude differences between the SDSS Petrosian and our measured 
Petrosian, isophotal magnitudes with surface brightness limits at 
25 ${\rm mag/arcsec^{2}}$  and 1\% of sky brightness respectively, as a function of the SDSS Petrosian half-light radius $r_{\rm 50, \,sdss7}$.
The red data points with error bars
are the median, lower (25 per cent) and upper (75 per cent) quartiles for
galaxies in bins of width 2 kpc in half-light radius except the
first bin which has a width of 4 kpc and the last bin which has a width
of 9 kpc to include all the remaining objects.
The medians shown in red points are 0.056, 0.093, 0.160, 0.234, 0.270 
mag in top panel, and 0.117, 0.125, 0.166, 0.205, 0.269 mag in middle panel, 
and 0.169, 0.189, 0.232, 0.277, 0.314 mag in bottom panel.}
\label{1053deltamag_logr50-fig10}
\end{figure}

Fig.~\ref{1053deltamag_logr50-fig10} shows the luminosity differences, $\dmpet$, $\dmiso$ and $\dmone$, 
as a function of $r_{\rm 50, \, sdss7}$ respectively.  
We can see from Fig.~\ref{1053deltamag_logr50-fig10} that 
there is a clear trend that $\dmone$ increases as ETGs become larger.

\begin{figure}
\centering
\includegraphics[width=1.0\textwidth]{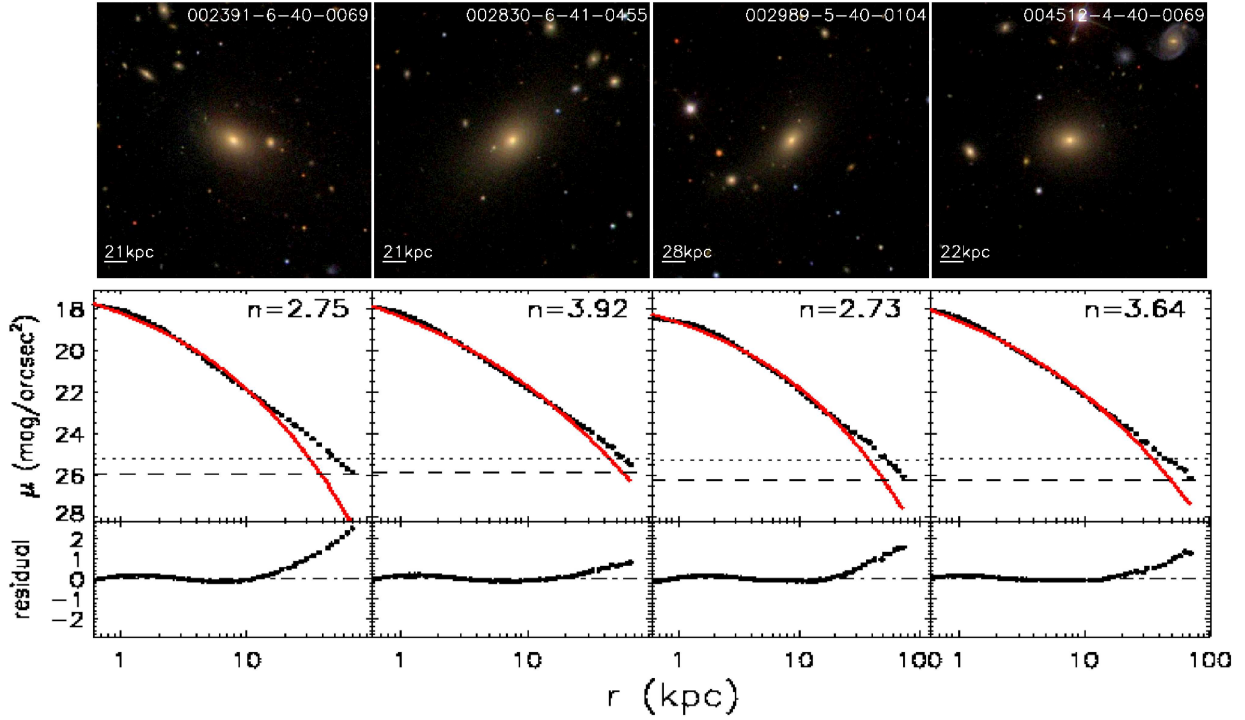}
\caption{Examples of bright ETGs with extended halos and corresponding 
surface brightness profiles. 
The red solid line shows the best fit S\' ersic law convolved with
the point spread function. The dotted and dashed lines in the middle panels represent the surface brightness of 
25 ${\rm mag/arcsec^{2}}$ and 1\% of sky brightness respectively.
The corresponding S\' ersic law index $n$ is shown in the top right of each panel.
The bottom panels show the residuals of the observed surface brightness from the 
 S\' ersic model.
}
\label{image2-fig11}
\end{figure}

We further examine the images of the brightest ETGs ($M_{\rm r}<-23$ mag) with
differences between the $r_{1\%}$ and the SDSS DR7 measurement larger than 10 kpc and $\dmone>0.4$ mag.  
In total, there are 102 such ETGs; 56\% ETGs are in crowded field, 26\% have extended stellar halos and
the remaining 18\% are contaminated by bright nearby stars.
Fig.~\ref{image2-fig11} represents the ETGs with extended stellar halos (top
panels) and corresponding surface brightness profiles (middle panels) and the residuals 
from S\' ersic models (bottom panels). 
It is obvious from Fig.~\ref{image2-fig11} that the bright ETGs with such very extended stellar halos 
could not be fitted well by a single S\' ersic model.

\begin{figure}
\centering
\includegraphics[width=1.0\textwidth]{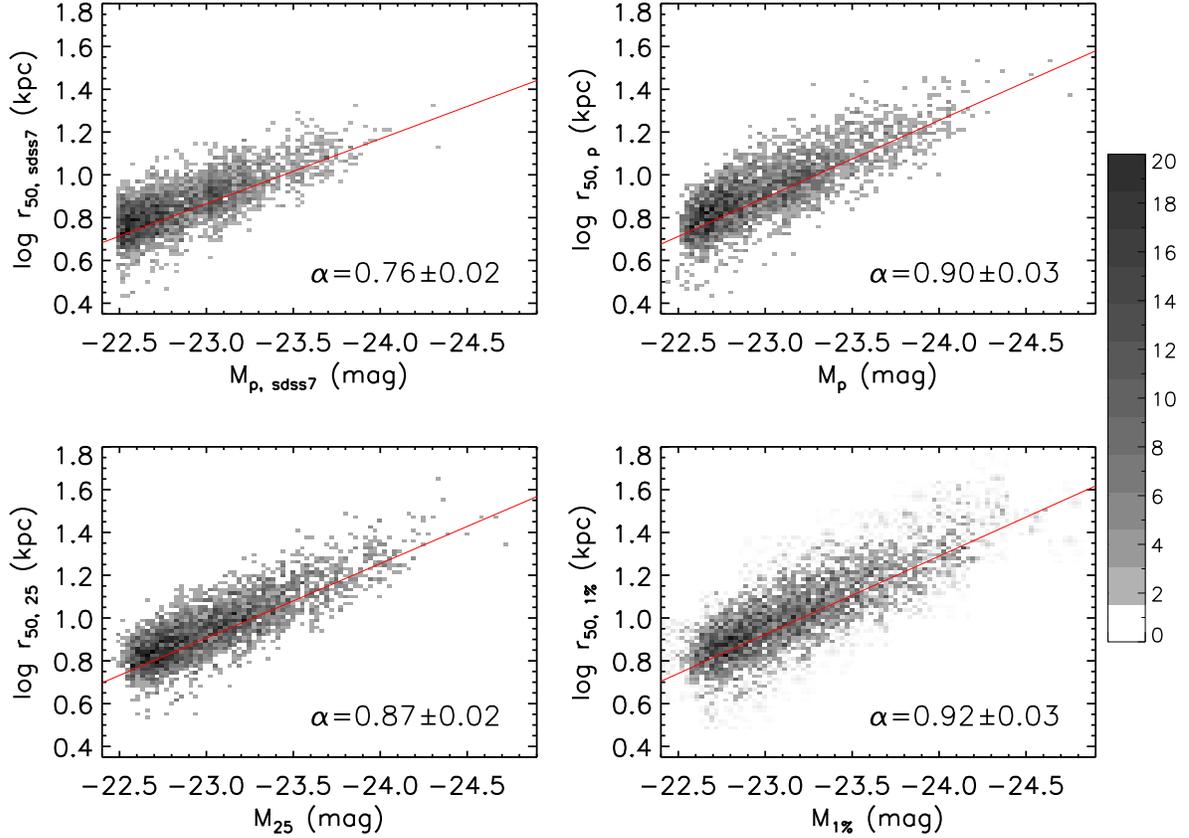}
\caption{The grayscale representation of the size-luminosity relation
  for 2949 bright ETGs. The four absolute magnitudes used are 
the SDSS Petrosian magnitude (top left), our own
Petrosian magnitude (top right), and aperture magnitudes to 25 ${\rm mag/arcsec^{2}}$ (bottom left) 
and 1\% of sky brightness (bottom right), respectively. The red line in each panel is the best power-law fit for the size-luminosity relation. 
The power-law index $\alpha$ ($r_{50} \propto L^\alpha$) is shown in the bottom right of each panel.
The rightmost vertical grayscale bar reflects the corresponding number of galaxies.
}

\label{2949logsize-luminosity-fig12}
\end{figure}

Fig.~\ref{2949logsize-luminosity-fig12} shows the size and luminosity relations
measured by different methods. In each panel, we give the best power-law fit
slope for the correlations. The top left and right panels are for Petrosian
half-light radius with Petrosian absolute magnitude obtained from the SDSS DR7
catalog and measured by us,
respectively.  The slope in the top right panel ($\alpha=0.90\pm 0.03$) is steeper than that in
the top left panel ($\alpha=0.76\pm 0.02$).  It indicates that the sky background
subtraction for bright ETGs significantly influences the size-luminosity relation.
The bottom left and right panels of Fig.~\ref{2949logsize-luminosity-fig12}
show the correlations between the size and isophotal absolute magnitude to 25
$\magsec$ and to 1\% of sky brightness, respectively.  We can see from the
bottom two panels of Fig.~\ref{2949logsize-luminosity-fig12} that the slope for
the correlation between $\log r_{50, 1\%}$ and $M_{1\%}$ ($\alpha=0.92\pm
0.03$) is somewhat steeper than the correlation between $\log r_{50, 25}$ and
$M_{25}$ ($\alpha=0.87\pm 0.02$). These results are consistent with the
size-luminosity relation trend for the brightest cluster galaxies (BCGs) obtained by
Liu et al. (2008) who found that the power-law slope becomes steeper
when the measurement goes deeper (see also Bernardi et al. 2007).

Our derived slope on the $\log r_{50}$ and $M_{\rm r}$ relation for ETGs 
is much larger than the value 0.67 found by Shen at al. (2003) that 
is widely adopted by recent works on the size evolution for
bright ETGs.  If we use our measured sizes of ETGs, the size evolution since redshift 2
will be somewhat larger. However, due to the surface brightness dimming, it may be difficult to perform  
photometry down to 26 $\magsec$ in the rest-frame $r$-band for ETGs at
redshift 2. As a result, we should be more cautious in discussing the size
evolution by explicitly taking into account the survey surface brightness limit.

\subsection{The bright end of the $r$-band LF and stellar mass density}

A basic way to investigate galaxy properties and their
evolution is by studying the luminosity function
(LF).  There are already many LF studies using different
samples and approaches at different redshifts (see the recent
review paper by Johnston 2011). Given that the luminosities of the brightest
galaxies ($-$23.5 mag) have been underestimated (by $\sim$10\% to 40\%), 
it is worth revisiting the LF at the bright end in the local universe.  

In this work, we construct the galaxy luminosity function at the bright end, 
utilizing the non-parametric $1/V_{\rm max}$ method (Schmidt 1968; Felten 1976; Eales 1993).
Our bright ETG sample includes 2949 galaxies consisting
of three subsamples as described in \S2, in which we have already discussed the
homogeneity for this sample. Briefly,
the $1/V_{\rm max}$ is the inverse of the maximum volume, 
to which the galaxy could
have been detected. The LF is obtained by integrating $1/V_{\rm max}$ in different
luminosity bins for the whole sample of galaxies. Given that our  ETGs
contain three subsamples, we calculate the
LF in three volumes separately, and then average them to obtain the final LF.
In addition, SDSS fiber collisions lead to $\sim 7\%$ incompleteness for the
spectroscopic sample (Bernardi et al. 2010), we multiply the $1/V_{\rm max}$ counts
by a factor of $1/0.93$ to obtain the final LF.

\begin{figure}
\centering
\includegraphics[width=0.85\textwidth]{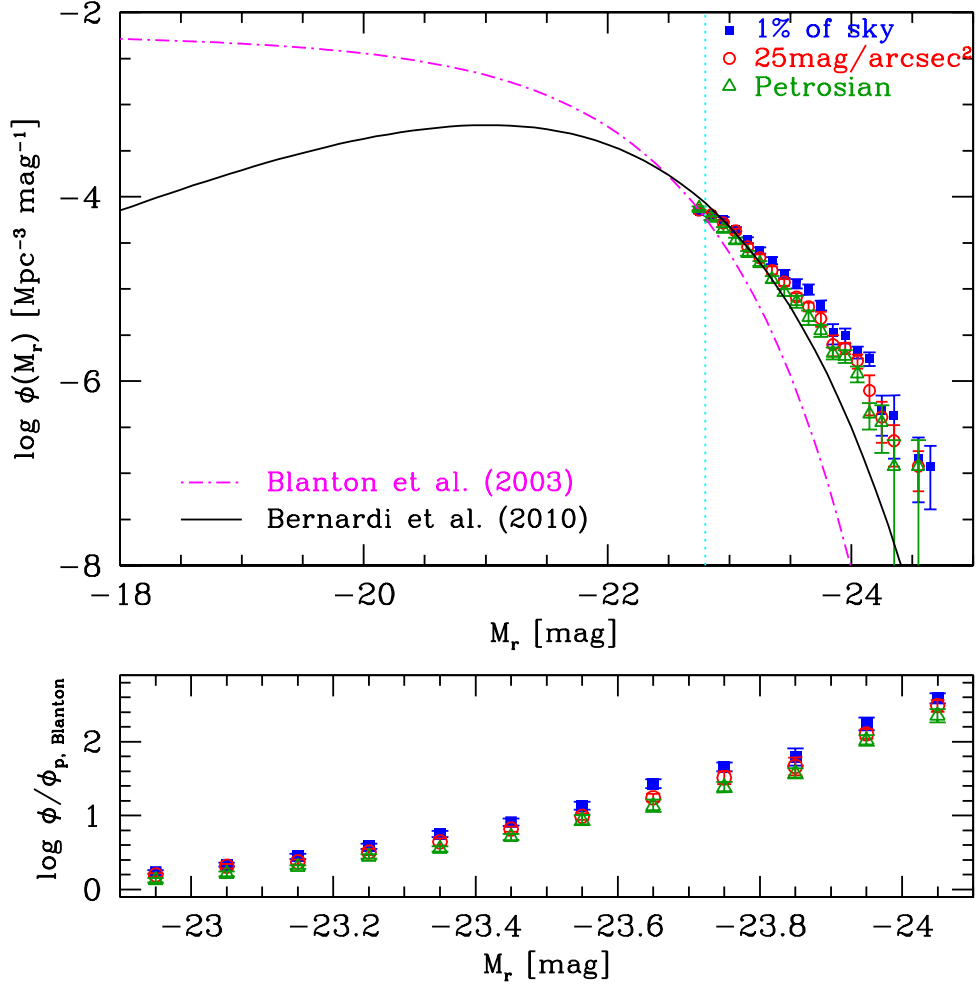}
\caption{The top panel shows the $r$-band luminosity function at the bright end calculated based
on our measured luminosities for 2949 ETGs brighter than $M_{\rm r}<-22.5$ mag.
The green triangles, red circles, and blue solid squares represent the LF calculated
by using our measured Petrosian magnitude, isophotal magnitudes with surface brightness limits at
 25 mag/arcsec$^2$ and 1\% of
sky brightness respectively. The magenta dot-dashed and black solid
lines give the fits by Blanton et al. (2003) and Bernardi et al. (2010).
The points for objects brighter than the luminosity shown by the cyan dotted line are not
affected by photometry.
The bottom panel shows the galaxy luminosity density ratios as a function of the $r$-band luminosity.
The green triangles, red circles, and blue solid squares show the ratios
for our measured Petrosian, isophotal luminosities to 25 mag/arcsec$^2$ and 1\% of sky brightness to that from Blanton et al. (2003).
}
\label{2949ETGs-LF-fig13}
\end{figure}

The top panel of Fig.~\ref{2949ETGs-LF-fig13} shows our $r$-band luminosity
function at the bright end. The green triangles, red circles and blue solid squares
represent the luminosity function for our measured Petrosian magnitude, 
isophotal magnitudes measured to the surface brightness 
of 25 $\magsec$ and 1\% of the sky brightness, respectively.  
For comparison, we also plot the luminosity
functions from Blanton et al. (2003) and Bernardi et al. (2010) in the top
panel of Fig.~\ref{2949ETGs-LF-fig13}. Blanton et al. (2003) used a
sample of 147,986 galaxies (from SDSS EDR) and
the maximum likelihood method to calculate
the luminosity function at $z=0.1$. Their sample is much larger, but includes all morphological types. However, at the
bright end  (e.g. the galaxies brighter than $-$22.5 mag), their galaxies
are almost exclusively ETGs.  The Petrosian $r$-band luminosity of galaxies 
in Blanton et al. (2003) are directly
obtained from the SDSS catalogue based on the SDSS PHOTO algorithm. Bernardi et al.
(2010), on the other hand, used an ETG sample of galaxies selected from $\sim 250000$ SDSS galaxies
with $14.5<m_{\rm p, \,sdss}<17.5$ using the concentration index $C_{r}\ge2.86$,
which is a conservative way to select ETGs from SDSS. 
The luminosities of ETGs in
Bernardi et al. (2010) are calculated using the cmodel in the SDSS pipeline 
with their own sky background subtraction method.

It is clear from the top panel of Fig.~\ref{2949ETGs-LF-fig13} that at the bright
end  ($M_{\rm r}< -22.5$ mag), the slope of the Bernardi et al. (2010)'s LF is
shallower than that of Blanton et al. (2003), implying that the luminosity
estimation method of Bernardi et al. (2010) is a 
significant improvement over the SDSS PHOTO algorithm. 
However, our luminosity function of ETGs at the bright end 
is even shallower, particularly when we use the photometry to 1\% of the sky brightness.

Table 1 lists our measured luminosity densities for the Petrosian magnitudes and isophotal magnitudes 
to surface brightness of 25 ${\rm mag/arcsec^2}$ and 1\% of the sky background (denoted by $\phi_{\rm p}$,
$\phi_{\rm 25}$ and $\phi_{\rm 1\%}$, respectively) at several $r$-band luminosities. The
bottom panel of Fig.~\ref{2949ETGs-LF-fig13} shows the ratios of these galaxy luminosity
densities to that measured by Blanton et al. (2003), denoted as $\phiPB$, 
 as a function of $r$-band luminosities. It is clear that the luminosity density ratios increase with the luminosity of ETGs. For ETGs with
 $M_{\rm r}=-$23.5 mag, the ratios are $6.7\pm0.46$, $8.0\pm0.40$ and $10.5\pm0.56$,
respectively; for ETGs with $M_{\rm r}=-$24 mag, they go
up to $153\pm14$, $197\pm16$ and $259\pm22$, respectively. It demonstrates that the luminosity
density of the brightest ETG calculated based on the SDSS catalogue has been
seriously underestimated. 
Note that for the three magnitudes $M_{\rm p}$, $M_{\rm 25}$ and $M_{\rm 1\%}$,
 the number of ETGs brighter than $-$23.5 mag is 347, 385 and 486, 
while the number of ETGs brighter than $-$24 mag is 41, 56 and 90,
and so the number statistics are reasonably good.
The underestimates in the luminosity density result in a significant underestimate of the integrated 
luminosity density. In particular, the integrated luminosity density down to $M_{\rm r}<-22.5$ mag based 
on our measured $M_{\rm p}$, $M_{25}$, $M_{1\%}$ are about 20\%, 40\%, 50\% higher than that from the SDSS Petrosian luminosity, respectively.

Next we estimate the stellar mass $M_{\ast}$ of the ETGs following Bernardi et al. (2010), utilizing the equation 
\begin{equation}
\log \,M_{\ast}/M_{\odot} =
1.097(g-r)-0.406-0.4(M_{\rm r}-4.67)-0.19z,
\end{equation}
where $M_{\ast}$ is the stellar mass (in solar units), $g-r$ is the
rest-frame color, $M_{\rm r}$ is the absolute magnitude, and $z$ is the
redshift.  This equation has already taken into account the k-correction and
evolution correction, and the initial mass function (IMF) is assumed to be of
the Chabrier (2003) form (see \S2.4 of Bernardi et al. 2010 for more details).
The first two terms on the right side of eq. (3) were initially derived by Bell
et al. (2003) based on SDSS Petrosian $g-r$ color and $r$-band Petrosian
magnigtude, which have been adapted to the Chabrier IMF here. A proper model fitting to colors and
magnitudes measured using our photometric methods may give different
coefficients in eq. (3).  However, to facilitate comparison with the stellar mass function obtained by
Bernardi et al. (2010), we use the same equation as their study. Under the
assumption that the impact of our photometric algorithm on the $r$-band
photometry is the same as that on the $g$-band, we use the SDSS model $g-r$ color as
a surrogate of colors measured by our methods\footnote{In the SDSS
Data Release 2 paper (Abazajian et al. 2004) and SDSS web page, model magnitudes are recommended
to be used for the measures of the colors of extended objects.}. The stellar masses $M_{\ast,
\rm \,p}$, $M_{\ast, \,25}$ and $M_{\ast, \,1\%}$ are obtained using our
measured luminosities $M_{\rm p}$, $M_{\rm 25}$, $M_{\rm 1\%}$, respectively. 

\begin{figure}
\centering
\includegraphics[width=0.85\textwidth]{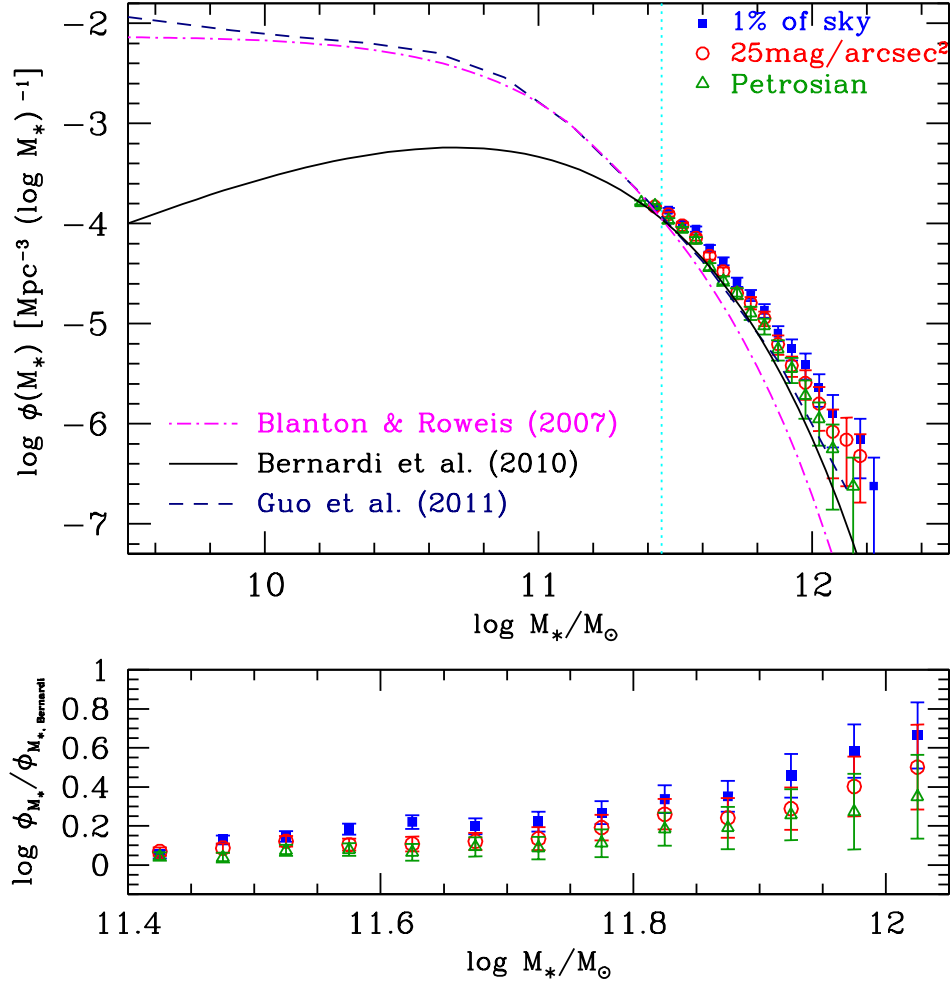}
\caption{ The top panel shows the stellar mass function for massive ETGs.
The stellar masses are estimated from our measured Petrosian
(green triangles), isophotal luminosities to 25 mag/arcsec$^2$
(red circles) and 1\% of sky brightness (blue solid squares), respectively.
The magenta dot-dashed and black solid lines represent the best-fit 
stellar mass functions from
Blanton \& Roweis (2007) and Bernardi et al. (2010), respectively.
The navy blue dashed line represents the prediction from 
the semi-analytic models of the Millennium Simulation of Guo et al. (2011). 
The points for objects more massive than the stellar mass shown by the cyan dotted line are not
affected by the photometry.
The bottom panel shows the ratios between our measured stellar mass
densities and those of Bernardi et al. (2010) as a function of stellar mass;
the symbols are the same as in the top panel.
}
\label{2949ETG-MF-fig14}
\end{figure}

Table 2 gives the stellar mass densities, $\phi_{M_\ast, \rm \,p}$, $\phi_{M_\ast, \,25}$ and 
$\phi_{M_\ast, \,1\%}$ at several stellar masses of ETGs for our measured magnitudes.
The top panel of Fig.~\ref{2949ETG-MF-fig14} shows the stellar mass function
for massive ETGs. We also plot the stellar mass function
from Blanton \& Roweis (2007) and Bernardi et al. (2010) in the top panel of 
Fig.~\ref{2949ETG-MF-fig14} for comparison. As can be seen, 
 the slope of our measured stellar mass functions is shallower than those 
of Bernardi et al. (2010) and Blanton \& Roweis (2007). 
Given that Bernardi et al. (2010) have already compared their stellar mass density with previous 
observational results, in the bottom panel of Fig.~\ref{2949ETG-MF-fig14} we just compare our result with Bernardi et al. (2010).  
It shows that the stellar mass densities ratios of our measurements to that of Bernardi et al. (2010), written as $\phi_{M_*, \rm \,Bernardi}$, as a function of stellar mass.  
We can see from the bottom panel of Fig.~\ref{2949ETG-MF-fig14} that 
the ratios are larger for more massive ETGs. 
The ratios, $\phi_{M_*, \rm \,p}/\phi_{M_*, \rm \,Bernardi}$, 
$\phi_{M_*, \,25}/\phi_{M_*, \rm \,Bernardi}$ and $\phi_{M_*, \,1\%}/\phi_{M_*, \rm \,Bernardi}$ 
are 1.2$\pm$0.06, 1.3$\pm$0.07 and 1.6$\pm$0.07 for ETGs with
$M_{\ast}\sim 5\times10^{11} M_{\odot}$; for $M_{\ast}\sim 10^{12}
M_{\odot}$, the ratios are even larger, 2.1$\pm$0.40, 2.8$\pm$0.43 and
4.2$\pm$0.58, respectively. Thus, at the high mass end the stellar mass densities have been underestimated by all previous works.

We also plot the predicted stellar mass function by Guo et al. (2011) in the top panel of
Fig.~\ref{2949ETG-MF-fig14}. Their result is based on the semi-analytic models of the galaxy population using the dark matter only Millennium Simulation. We can see that the high-mass tail of 
their stellar mass function slightly over-predicts the abundance in Bernardi et al. (2010), but still 
under-predicts our mass function. We return to this briefly in the next section.

\section{SUMMARY}

In this work, we performed photometric analyses for a complete and homogeneous
bright ETGs sample with 2949 early-type galaxies ($M_{\rm r}<-22.5$ mag) in the
redshift range of 0.05 to 0.15, taken from the catalog of SDSS DR7; all these 
galaxies have morphological classifications from the Galaxy Zoo 1 MGS.
Based on our own sky background subtraction method, we 
measured the Petrosian and isophotal magnitudes, as well as
the corresponding half-light radii. Comparing our measured luminosities and
sizes to those from SDSS, we find that the SDSS pipeline
significantly underestimates the luminosities and sizes for the brightest ETGs,
leading to underestimates of the luminosity density and stellar
mass density for bright ETGs. Our main results are summarized as
follows.

\begin{enumerate}
\item We find that for brightest galaxies ($M_{\rm}<-23$ mag), 
our Petrosian magnitudes, and isophotal magnitudes to 
25 ${\rm mag/arcsec^2}$ and 1\% of the sky brightness are on
average 0.16 mag, 0.20 mag, and 0.26 mag brighter than the SDSS Petrosian values, respectively.
In the first case the underestimations are due to overestimations in the sky
background by the SDSS PHOTO algorithm, while the latter two are also caused by
an additional effect of reaching deeper photometry.
Such underestimation is more severe as ETGs become more luminous. Our results also demonstrate that as 
we integrate to deeper surface brightness, we recover more luminosity of galaxies.
\item We also find that the sizes of ETGs (half-light radius $r_{50}$) measured by the SDSS algorithm are smaller than those measured by us. 
The largest $r_{50}$ of bright ETGs in the SDSS catalogue is $\sim$ 20 kpc, while our measured $r_{50}$ can be as large as
30 kpc and for a large fraction ($\sim$ 27\%) of brightest ETGs, $r_{50}$ is larger than 15 kpc. 
In addition, we find that the slope in the size-luminosity relation at the bright end is much steeper than that
found by Shen et al. (2003). 
\item Based on our selected complete and homogeneous sample of 2949 bright early-type galaxies,
we construct the luminosity function. We find that the LF at the bright end is
much shallower than those of Blanton et al. (2003) and Bernardi et al. (2010). 
The luminosity density at $-$23.5 mag ($-$24 mag) measured in this work is one order (two orders) of magnitude higher than 
that of Blanton et al. (2003). 
As a result, the ratios of the integrated luminosity density for bright 
galaxies ($M_{\rm r}<-22.5$ mag) between those based on our measured 
$M_{\rm p}$, $M_{25}$, $M_{1\%}$ and the SDSS Petrosian luminosity are 1.2, 1.4 and 1.5, respectively.
Similarly, the stellar mass density of ETGs is a few tenths to a factor of few higher than that 
of Bernardi et al. (2010) for stellar mass from $\sim 5\times10^{11} M_{\odot}$ to $\sim 10^{12} M_{\odot}$.
Therefore, our method recovers substantially more luminosity for bright ETGs, which 
may alleviate the contradiction between hierarchical galaxy formation theories and current observations.
\end{enumerate}
Our results suggest that very careful photometry needs to be performed to
obtain the LF at the bright end which has significant impact on the
stellar mass function for massive galaxies. Previous claims that the
massive end of the stellar mass function has a weak evolution since
redshift $\sim$ 1 in comparison with the Galaxy And Mass Assembly (GAMA)
survey (Baldry et al. 2012) may need to be re-examined. Furthermore,
semi-analytical models may need to increase their cooling efficiency or
decrease the AGN feedback efficiency for massive galaxies in order to
match the shallower LF we found here. How does this affect the overall
shape of the LF (especially around $L_*$) is unclear and warrants
further studies. 
Finally, we notice that recent studies on the environmental
dependence of galaxy mass function have suggested that the stellar mass 
function is much more dependent on local galaxy density than global 
environments of galaxies and the most massive galaxies are only located in the highest density regions (e.g. Vulcani et al. 2012; Calvi et al. 2013). 
Given our finding that massive galaxies, which
are preferentially found in crowded environments, can be significantly
underestimated in their luminosity/mass due to inaccurate photometry,
the observed environmental dependence of the high-mass end of mass
function may be somehow affected by such photometry issue also.
It is worth revisiting this issue in a future work.

\acknowledgements
We thank Drs. J. S. Huang, F. S. Liu, Z. H. Shang, Qi Guo, H. J. Yan and C. G. Shu for advice and helpful discussions.
This project is supported by the NSF of China 10973011, 10833006, 11003015, the Chinese Academy of Sciences and NAOC (SM).
YPJ is supported by NSFC (11033006, 11121062) and the CAS/SAFEA International Partnership Program for Creative Research Teams (KJCX2-YW-T23).
Funding for the creation and distribution of the SDSS Archive has been provided
by the Alfred P. Sloan Foundation, the Participating Institutions, the National
Aeronautics and Space Administration, the National Science Foundation, the U.S. Department of Energy,
the Japanese Monbukagakusho, and the Max Planck Society. The SDSS Web site is http://www.sdss.org/.
The SDSS is managed by the Astrophysical Research Consortium (ARC) for the
Participating Institutions. The Participating Institutions are The University
of Chicago, Fermilab, the Institute for Advanced Study, the Japan Participation
Group, The Johns Hopkins University, the Korean Scientist Group, Los Alamos
National Laboratory, the Max-Planck-Institute for Astronomy (MPIA), the
Max-Planck-Institute for Astrophysics (MPA), New Mexico State University,
University of Pittsburgh, Princeton University, the United States Naval Observatory,
and the University of Washington.

\acknowledgments

\email{}

\clearpage
\pagestyle{empty}
\begin{deluxetable}{l r r r}  
\tablenum{1}
\tablewidth{0pt}
\tabletypesize{\tiny}
\tablecaption{Luminosity densities for the Petrosian magnitude
  ($\phi_{\rm p}$) and isophotal magnitudes to 25 mag/arcsec$^2$
  ($\phi_{\rm 25}$) and 1\% of the sky background ($\phi_{\rm 1\%}$).}
\tablehead{
\multicolumn{1}{c}{$M_{\rm r}$} &
\multicolumn{1}{c}{$\phi_{\rm p}$} &
\multicolumn{1}{c}{$\phi_{\rm 25}$} &
\multicolumn{1}{c}{$\phi_{\rm 1\%}$} \\
\multicolumn{1}{c}{(mag)} &
\multicolumn{1}{c}{($10^{-7}$Mpc$^{-3}$ mag$^{-1}$)} &
\multicolumn{1}{c}{($10^{-7}$Mpc$^{-3}$ mag$^{-1}$)} &
\multicolumn{1}{c}{($10^{-7}$Mpc$^{-3}$ mag$^{-1}$)} 
}
\startdata
-22.75\tablenotemark{a} &  751.745$\pm$34.527  & 716.908$\pm$33.718  & 691.239$\pm$33.109 \\
-22.85 &  588.563$\pm$30.551  & 634.400$\pm$31.718  & 645.401$\pm$31.992 \\
-22.95 &  443.715$\pm$26.526  & 522.556$\pm$28.787  & 573.894$\pm$30.168 \\
-23.05 &  335.537$\pm$23.067  & 425.380$\pm$25.973  & 445.300$\pm$26.251 \\
-23.15 &  244.257$\pm$13.778  & 286.031$\pm$21.298  & 340.198$\pm$24.652 \\
-23.25 &  189.786$\pm$12.098  & 218.191$\pm$18.601  & 262.195$\pm$20.391 \\
-23.35 &  127.568$\pm$9.925  & 159.517$\pm$15.905  & 203.522$\pm$17.965 \\
-23.45 &  92.339$\pm$8.424  & 118.448$\pm$9.555  & 146.683$\pm$15.252 \\
-23.55 &  73.341$\pm$10.785  & 82.101$\pm$7.963  & 114.717$\pm$13.132 \\
-23.65 &  49.505$\pm$8.860  & 64.004$\pm$7.017  & 99.011$\pm$12.530 \\
-23.75 &  35.526$\pm$5.226  & 47.672$\pm$8.695  & 66.697$\pm$8.242 \\
-23.85 &  20.151$\pm$2.954  & 24.958$\pm$6.438  & 33.346$\pm$8.371 \\
-23.95 &  18.670$\pm$2.945  & 22.846$\pm$3.195  & 31.383$\pm$5.785 \\
-24.05 &  12.094$\pm$2.348  & 16.537$\pm$2.857  & 20.738$\pm$3.073 \\
-24.15 &  4.397$\pm$1.394  & 7.915$\pm$3.681  & 17.561$\pm$3.025 \\
-24.25 &  3.582$\pm$1.923  & 4.075$\pm$1.923  & 4.776$\pm$2.221 \\
-24.35 &  1.194$\pm$1.111  & 2.262$\pm$1.071  & 4.234$\pm$2.784 \\
\enddata
\tablenotetext{a}{The luminosity density at $M_{r}=-22.75$ mag is affected by the photometry.}
\label{tab:lf}          
\end{deluxetable}

\clearpage

\clearpage
\pagestyle{empty}
\begin{deluxetable}{l r r r}
\tablenum{2}
\tablewidth{0pt}
\tabletypesize{\tiny}
\tablecaption{Stellar mass densities for the Petrosian magnitude
  ($\phi_{M_\ast, \rm \,p}$) and isophotal magnitudes to 25
  mag/arcsec$^2$ ($\phi_{M_\ast, \rm \,25}$) and 1\% of the sky
  background ($\phi_{M_\ast, \rm \,1\%}$).}
\tablehead{
\multicolumn{1}{c}{$\log \, M_\ast/M_\odot$} &
\multicolumn{1}{c}{$\phi_{M_\ast, \rm \,p}$} &
\multicolumn{1}{c}{$\phi_{M_\ast, \rm \,25}$} &
\multicolumn{1}{c}{$\phi_{M_\ast, \rm \,1\%}$} \\
\multicolumn{1}{c}{} &
\multicolumn{1}{c}{($10^{-7}$ Mpc$^{-3}\, {\rm dex}^{-1}$)} &
\multicolumn{1}{c}{($10^{-7}$ Mpc$^{-3}\,  {\rm dex}^{-1}$)} &
\multicolumn{1}{c}{($10^{-7}$ Mpc$^{-3}\,  {\rm dex}^{-1}$)} 
}
\startdata
11.425\tablenotemark{a} &  1503.490$\pm$69.055  & 1481.490$\pm$68.547  & 1437.480$\pm$67.522 \\
11.475 &  1078.120$\pm$58.476  & 1239.460$\pm$62.699  & 1364.140$\pm$65.777 \\
11.525 &  876.429$\pm$52.723  & 964.438$\pm$55.307  & 982.773$\pm$55.830 \\
11.575 &  685.742$\pm$46.636  & 722.413$\pm$47.867  & 880.096$\pm$52.833 \\
11.625 &  370.374$\pm$34.274  & 476.719$\pm$38.884  & 568.396$\pm$42.459 \\
11.675 &  267.674$\pm$28.369  & 333.723$\pm$31.676  & 421.713$\pm$36.572 \\
11.725 &  202.038$\pm$24.976  & 201.688$\pm$25.292  & 260.721$\pm$27.998 \\
11.775 &  128.623$\pm$19.665  & 161.351$\pm$22.622  & 192.372$\pm$24.181 \\
11.825 &  95.344$\pm$17.390  & 113.685$\pm$18.989  & 135.768$\pm$20.758 \\
11.875 &  55.620$\pm$12.932  & 62.573$\pm$13.716  & 80.665$\pm$13.755 \\
11.925 &  35.519$\pm$9.903  & 38.211$\pm$8.884  & 56.210$\pm$13.499 \\
11.975 &  19.155$\pm$7.968  & 25.752$\pm$8.448  & 39.100$\pm$11.500 \\
12.025 &  11.204$\pm$5.141  & 15.896$\pm$7.393  & 23.064$\pm$8.333 \\
12.075 &  5.627$\pm$4.233  & 8.355$\pm$5.495  & 12.604$\pm$6.768 \\
\enddata
\tablenotetext{a}{The stellar mass density at $\log \, M_\ast/M_\odot=11.425$ is affected by the photometry.}
\label{tab:mf}
\end{deluxetable}

\clearpage  



\begin{references}
\reference{}Abazajian K., Adelman-McCarthy, J. K., Ag\" ueros, M. A., et al. 2004, AJ, 128, 502
\reference{}Aihara, H., Allende, P. C., An, D., et al. 2011, ApJS, 193, 29
\reference{}Baldry, I. K., Driver, S. P., Loveday, J., et al. 2012, MNRAS, 421, 621
\reference{}Bell, E. F., McIntosh, D. H., Katz, N., Weinberg, M. D. 2003, ApJS, 149, 289 
\reference{}Bernardi, M., Hyde, J. B., Sheth, R. K., Miller, C. J., Nichol, R. C. 2007, AJ, 133, 1741
\reference{}Bernardi, M., Shankar, F., Hyde, J. B., et al. 2010, MNRAS, 404, 2087 
\reference{}Bertin, E. \& Arnouts, S. 1996, A\&A, 117, 393 
\reference{}Bezanson, R., van Dokkum, P. G., Tal, T., et al. 2009, ApJ, 697, 1290 
\reference{}Blanton, M. R., Hogg, D. W., Bahcall, N. A., et al. 2003, ApJ, 592, 819 
\reference{}Blanton, M. R. \& Roweis, S. 2007, AJ, 133, 734 
\reference{}Brough, S., Tran, K.-V, Sharp, R. G., von der Linden, A., Couch, W. J., 2011, MNRAS, 414, L80
\reference{}Bruce, V. A., Dunlop, J. S., Cirasuolo, M., et al. 2012, MNRAS, 427, 1666 
\reference{}Calvi, R., Poggianti, B. M., Vulcani, B., Fasano, G., 2013, MNRAS, in press (arXiv:1304.3124)
\reference{}Cappellari, M., McDermid, R. M., Alatalo, K., et al. 2012, MNRAS, in press (arXiv:1208.3523)
\reference{}Chabrier G., 2003, ApJ, 586, L133 
\reference{}Cimatti, A., Daddi, E., Renzini, A. 2006, A\&A, 453, L29
\reference{}Cimatti, A., Cassata, P., Pozzetti, L., et al. 2008, A\&A, 482, 21
\reference{}Cimatti, A. 2009, AIPC, 1111, 191
\reference{}Cool, R. J., Eisenstein, D. J., Fan, X. H., Fukugita, M., Jiang, L. H., et al. 2008, ApJ, 682, 919
\reference{}Cowie, L. L., Songaila, A., Hu, E. M., Cohen, J. G. 1996, AJ, 112, 839
\reference{}Daddi, E., Renzini, A., Pirzkal, N., et al. 2005, ApJ, 626, 680
\reference{}De Lucia, G. \& Blaizot, J. 2007, MNRAS, 375, 2
\reference{}De Lucia, G. \& Borgani, S. 2012, MNRAS, 426, L61
\reference{}Eales, S. 1993, ApJ, 404, 51
\reference{}Emsellem, E., Cappellari, M., Krajnovi\' c, D., et al. 2007, MNRAS, 379, 401
\reference{}Felten, J. E. 1976, ApJ, 207, 700
\reference{}Fontanot, F., De Lucia, G., Monaco, P., Somerville, R. S., Santini, P. 2009, MNRAS, 397, 1776
\reference{}Fukugita, M., Nakamura, O., Okamura, S., et al. 2007, AJ, 134, 579 
\reference{}Gavazzi, G., \& Scodeggio, M. 1996, A\&A, 312, L29
\reference{}Graham, A. W., Driver, S. P., Petrosian, V., et al. 2005, AJ, 130, 1535
\reference{}Guo, Q., White, S., Bryna-KAlwin, M., et al. 2011, MNRAS, 413, 101
\reference{}Johnston, R. 2011, A\&ARv, 19, 41
\reference{}Kaviraj, S., Schawinski, K., Devriendt, J. E. G., et al. 2007, ApJS, 173, 619 
\reference{}Lintott, C., Schawinski, K., Bamford, S., et al. 2011, MNRAS, 410, 166 
\reference{}Liu, F. S., Xia, X. Y., Mao, S. D., Wu, H., Deng, Z. G. 2008, MNRAS, 385, 23
\reference{}McLure, R. J., Pearce, H. J., Dunlop, J. S., et al. 2012, MNRAS, in press (arXiv:1205.4058)
\reference{}Naab T. 2012, in Proc. of the XXVIII IAU General Assembly, eds. D. Thomas, A. Pasquali; I. Ferreras. Cambridge University Press, in press (arXiv:1211.6892)
\reference{}Petrosian, V. 1976, ApJ, 209, L1 
\reference{}Poggianti, B. M., Calvi, R., Bindoni, D., et al. 2013, ApJ, 762, 77 
\reference{}Renzini, A. 2006, ARA\&A, 44, 141
\reference{}Scarlata, C., Carollo, C. M., Lilly, S. J., et al. 2007, ApJS, 172, 494
\reference{}Schawinski, K., Kaviraj, S., Khochfar, S., et al. 2007, ApJS, 173, 512 
\reference{}Schmidt, M. 1968, ApJ, 151, 393 
\reference{}S\' ersic, J. L. 1963, Bol. Asoc. Argent. Astron., 6, 41
\reference{}Shen, S., Mo, H. J., White, S. D. M., et al. 2003, MNRAS, 343, 978
\reference{}Stoughton, C., Lupton, R. H., Bernardi, M., et al. 2002, AJ, 123, 485
\reference{}Strauss, M. A., Weinberg, D. H., Lupton, R. H., et al. 2002, AJ, 124, 1810
\reference{}Szomoru, D., Franx, M., van Dokkum, P. 2012, ApJ, 749, 121
\reference{}Tal, T. \& van Dokkum, P. G. 2011, ApJ, 731, 89
\reference{}Tonini, C., Bernyk, M., Croton, D., Maraston, C., Thomas, D. 2012, ApJ, 759, 43
\reference{}Trujillo, I., F\" orster, S., Natascha M., et al. 2006, ApJ, 650, 18
\reference{}Trujillo, I. 2012, in Proc. of the XXVIII IAU General Assembly, eds. D. Thomas, A. Pasquali; I. Ferreras. Cambridge University Press, in press (arXiv:1211.3771)
\reference{}Valentinuzzi, T., Fritz, J., Poggianti, B. M., et al. 2010, ApJ, 712, 226 
\reference{}van Dokkum, P. G., Franx, M., Kriek, M., et al. 2008, ApJ, 677, L5
\reference{}van Dokkum, P. G., Whitaker, K. E., Brammer, G., et al. 2010, ApJ, 709, 1018
\reference{}von der Linden, A., Best, P. N., Kauffmann, G., White, S. D. M. 2007, MNRAS, 379, 867
\reference{}Vulcani, B., Poggianti, B. M., Fasano, G., et al. 2012, MNRAS, 420, 1481 
\reference{}Vulcani, B., Poggianti, B. M., Arag\' on-Salamanca, A., et al. 2011, MNRAS, 412, 246 
\reference{}Wu, H., Shao, Z. Y., Mo, H. J., Xia, X. Y., Deng, Z. G. 2005, ApJ, 622, 244
\reference{}York, D. G., Adelman, J., Anderson, J. E., et al. 2000, AJ, 120, 1579

\end{references}
\end{document}